\documentclass[aps,prd,twocolumn,preprintnumbers,superscriptaddress,nofootinbib,floatfix]{revtex4-1}

\usepackage[dvipsnames]{xcolor}
\usepackage{amssymb,amsmath,latexsym,mathrsfs}
\usepackage[USenglish,american]{babel}
\usepackage{bm}
\usepackage{amsfonts}
\usepackage{graphicx}
\usepackage{epstopdf}
\usepackage{hyperref}
\usepackage{array}
\usepackage[utf8]{inputenc}
\usepackage{soul}
\usepackage{color}
\usepackage{slashed}
\usepackage{csquotes}
\usepackage[T1]{fontenc}

\enlargethispage{\baselineskip}

\definecolor{steelblue}{RGB}{25,25,112}
\hypersetup{colorlinks,linkcolor={blue},citecolor={blue},urlcolor={steelblue}}  

\everymath{\displaystyle} 
\newcommand*{\bfrac}[2]{\genfrac{}{}{0pt}{}{#1}{#2}}

\newcommand{\THi}{\tau_{\rm Higgs}}

\newcommand{\hh}{\langle h^2\rangle}

\newcommand{\WW}{\mathcal{W}_k}

\begin{document}
\title{Large Density Perturbations from Reheating to Standard Model particles due to the Dynamics of the Higgs Boson during Inflation
}

\newcommand{\FIRSTAFF}{\affiliation{The Oskar Klein Centre for Cosmoparticle Physics,
	Department of Physics,
	Stockholm University,
	AlbaNova,
	10691 Stockholm,
	Sweden}}
\newcommand{\SECONDAFF}{\affiliation{Department of Physics,
  University of Texas,
  Austin, TX 78722}}
\newcommand{\THIRDAFF}{\affiliation{Nordita,
	KTH Royal Institute of Technology and Stockholm University
	Roslagstullsbacken 23,
	10691 Stockholm,
	Sweden}}
\newcommand{\FOURTHAFF}{\affiliation{Institut de F{\'i}sica d'Altes Energies (IFAE), The Barcelona Institute of Science and Technology (BIST), Campus UAB, 08193 Bellaterra, Barcelona}}
\newcommand{\FIFTHAFF}{\affiliation{Nikhef,
	Science Park 105,
	1098 XG Amsterdam, The Netherlands}}	
\newcommand{\SIXTHAFF}{\affiliation{Institute Lorentz of Theoretical Physics,
	University of Leiden, 2333CA Leiden, The Netherlands}}
\newcommand{\SISSA}{\affiliation{Scuola Internazionale Superiore di Studi Avanzati (SISSA), via Bonomea 265, 34136 Trieste, Italy}}
\newcommand{\INFNTrieste}{\affiliation{INFN, Sezione di Trieste, via Valerio 2, 34127 Trieste, Italy}}
\newcommand{\IFPU}{\affiliation{Institute for Fundamental Physics of the Universe (IFPU), via Beirut 2, 34151 Trieste, Italy}}
\newcommand{\GRAPPA}{\affiliation{Gravitation Astroparticle Physics Amsterdam (GRAPPA),\\Institute for Theoretical Physics Amsterdam and Delta Institute for Theoretical Physics,\\ University of Amsterdam, Science Park 904, 1098 XH Amsterdam, The Netherlands}}

\author{Aliki Litsa}
\email[Electronic address: ]{aliki.litsa@fysik.su.se}
\FIRSTAFF

\author{Katherine Freese}
\email[Electronic address: ]{ktfreese@utexas.edu}
\FIRSTAFF
\SECONDAFF
\THIRDAFF

\author{Evangelos I. Sfakianakis}
\email[Electronic address: ]{e.sfakianakis@nikhef.nl}
\FOURTHAFF
\FIFTHAFF
\SIXTHAFF

\author{Patrick Stengel}
\email[Electronic address: ]{pstengel@sissa.it}
\FIRSTAFF
\SISSA
\INFNTrieste
\IFPU

\author{Luca Visinelli}
\email[Electronic address: ]{luca.visinelli@lnf.infn.it}
\GRAPPA
\affiliation{INFN, Laboratori Nazionali di Frascati, C.P. 13, 100044 Frascati, Italy}

\date{\today}
\preprint{UT-03-2020; UTTG-11-2020; Nikhef 2020-029; NORDITA-2020-082}

\begin{abstract}

Cosmic Microwave Background (CMB) observations are used to constrain reheating to Standard Model (SM) particles after a period of inflation. As a light spectator field, the SM Higgs boson acquires large field values from its quantum fluctuations during inflation, gives masses to SM particles that vary from one Hubble patch to another, and thereby produces large density fluctuations.  We consider both perturbative and resonant decay of the inflaton to SM particles. For the case of perturbative decay from coherent oscillations of the inflaton after high scale inflation, we find strong constraints on the reheat temperature for the inflaton decay into heavy SM particles. For the case of resonant particle production (preheating) to (Higgsed) SM gauge bosons, we find temperature fluctuations larger than observed in the CMB for a range of gauge coupling that includes those found in the SM and conclude that such preheating cannot be the main source of reheating the Universe after inflation.

\end{abstract}

\maketitle

\section{Introduction}
\label{sec:Introduction}

Inflation~\cite{Guth:1980zm, Sato:1980yn, Brout:1977ix} is a period of accelerated expansion that occurred in the very early epoch of our Universe. It was first proposed to explain the homogeneity, isotropy, and flatness observed in the cosmic microwave background (CMB) radiation~\cite{Guth:1982ec, Starobinsky:1982ee}, as well as the lack of relic monopoles. A mechanism for driving the dynamics of inflation comes in the form of a rolling scalar field, the inflaton~\cite{Linde:1981mu, Albrecht:1982wi}. In this framework, the density perturbations that are observed in the CMB are explained by the quantum fluctuations of the inflaton field. It is these perturbations that later develop into the large scale structure observed in the Universe~\cite{Colless:2001gk, 2017AJ....154...28B}. The most stringent constraints to the theory of inflation come from the observations of the CMB by the \textit{Planck} satellite, including the power spectrum~\cite{Akrami:2018odb} and bispectrum~\cite{Akrami:2019izv} of temperature anisotropies. 

The inflationary period must end by successfully reheating the Universe, which marks the transition into the radiation dominated cosmological era before Big Bang Nucleosynthesis (BBN) occurs. If inflation is driven by a rolling scalar field, reheating can occur through the decay of the inflaton into light degrees of freedom in the Standard Model (SM) or an intermediate sector. Typical mechanisms for reheating are perturbative decay of the inflaton~\cite{Dolgov:1982th, Abbott:1982hn} and resonant particle production~\cite{Chung:1999ve}. In particular, the inflaton field $\phi$ might have decayed due to the presence of interaction terms in the Lagrangian such as $\bar\psi\psi\phi$ and $\phi F^{\mu\nu} \tilde F_{\mu\nu}$. The former term is a Yukawa-type coupling to a fermion $\psi$ and the latter is a Chern-Simons coupling to gauge bosons with a field strength $F^{\mu\nu}$, as found in models of natural inflation~\cite{Freese:1990rb}. Other scenarios for reheating include the decay of the inflaton condensate into its own quanta, which must ultimately decay into SM particles, or through gravitational interactions~\cite{Starobinsky:1980te}. 

In this paper we consider reheating via inflaton decay to SM particles coupled to the Higgs boson. We note that the inflaton in this paper is not the Higgs boson; instead the Higgs is a light spectator that plays an important role in the reheating process. The scenario we consider is minimal in the sense that no new particles beyond the SM are introduced other than the inflaton itself.
Along with the inflaton flat direction which is the main component in driving the expansion rate of the Universe during the inflationary stage, the Higgs boson and other light fields that are present at this epoch would act as spectators since they would not directly affect the evolution of the background geometry. However, light spectator fields would acquire large quantum fluctuations and thereby effective masses that vary from one Hubble patch to another. If the light spectators are also associated with the decay of the inflaton field in each Hubble patch, their stochastic dynamics can cause spatial fluctuations in the reheat temperature and large density perturbations.

Inhomogeneous reheating due to the stochastic behavior of a light spectator field is known as {\it modulated reheating}~\cite{Dvali:2003em,Kofman:2003nx,Dvali:2003ar, Ichikawa:2008ne,Kobayashi:2011hp}. Examples of light spectators can include the SM Higgs boson, with mass $\sim \mathcal{O}\left(125\right)\,$GeV~\cite{Aad:2012tfa, Chatrchyan:2012ufa} and the hypothetical axion, with a mass typically well below the MeV range (see Ref.~\cite{DiLuzio:2020wdo} for a recent review). In this paper we focus on modulated reheating caused by coupling of the inflaton decay products to the SM Higgs boson~\cite{DeSimone:2012qr,Choi:2012cp,Cai:2013caa,Fujita:2016vfj}, which is taken to be a light spectator during inflation. We assume that the inflaton couples primarily to SM particles that develop masses when the Higgs field acquires a Vacuum Expectation Value (VEV). If the inflaton were instead to decay to a massless gauge mode in the broken phase (the analog of the photon at lower temperatures) or to un-Higgsed neutrinos, then the effective Higgs mass during inflation would be irrelevant to the reheating process, since the SM Higgs does not couple to these particles\footnote{We note that the direction of the Higgs VEV may or may not coincide with the direction of Spontaneous Symmetry Breaking at low temperatures, hence the massless direction during inflation is not necessarily the same as today's photon.}.  

Enqvist et al.~\cite{Enqvist:2013kaa} showed that during inflation, due to the quantum fluctuations of the Higgs field, the Higgs  can develop a mass so that electroweak (EW) symmetry can be treated as effectively broken~\cite{Kusenko:2014lra, Adshead:2015jza}. The expectation value of the Higgs amplitude over the entire inflating patch is vanishing $\langle h_I \rangle = 0$ due to the symmetric potential (where the subscript $I$ is used to indicate the initial value at the onset of inflaton oscillations, i.e. at the end of inflation). However, due to the quantum fluctuations, the variance is non-zero.  Thus, the  typical Higgs amplitude, i.e. the effective VEV, in a typical Hubble patch at the end of inflation is given by a root mean square value $h_I = \sqrt{\hh} \propto H_I$ where $H_I$ is the Hubble scale at the onset of inflaton oscillations. Even assuming that the energy density of the Higgs field is always subdominant to that of the inflaton, its effective VEV would be driven by the stochastic dynamics to relatively large values. The effective nonzero Higgs VEV during inflation then gives mass to all SM particles that couple to the Higgs. In turn, this would affect the decay of the inflaton to SM particles. Although usually considered to be much lighter than the inflaton, SM particles with masses due to the Higgs boson condensate which develops during inflation~\cite{Enqvist:2013kaa, Enqvist:2014bua} can have several interesting effects.  

In our previous paper (Ref.~\cite{Freese:2017ace} hereafter referred to as FSSV), we studied the effects of {\it Higgs blocking}, i.e. the delay in the reheating process due to the large particle masses acquired during inflation due to the effective Higgs VEV.  As long as the particle masses exceed the inflaton mass, reheating cannot occur. Only once the Higgs condensate decays do the particle masses vanish and reheating can proceed.  We studied Higgs blocking for the cases of both perturbative decay and resonant particle production. We also briefly discussed the potential for generating large temperature fluctuations due to the stochastic nature of the Higgs blocking, with variation from one Hubble patch to another. Subsequently, Ref.~\cite{Lu:2019tjj} calculated signatures for the effects of the Higgs on reheating in the Cosmological Collider framework and Ref.~\cite{Karam:2020skk} showed the potentially large temperature fluctuations which can arise due to Higgs blocking when the inflaton decays to fermions with relatively large Yukawa couplings to the SM Higgs. In our present work we extend our previous results of FSSV and treat both Higgs blocking and Higgs modulation to derive the corresponding density fluctuations that arise during reheating. For the case of perturbative inflaton decays, our previous work in FSSV found Higgs blocking to be negligible for SM fermions with Yukawa couplings $y < 1$;  here we will show that, even for this case, Higgs-modulation alone can indeed generate large temperature anisotropies. 

In this work we consider a simple reheating scenario where the inflaton (in the post-inflationary epoch) proceeds along a single field direction with a potential approximated by that of a massive scalar field. We do not further specify any particular inflationary potential. We must again stress that the inflaton field in this study is not the Higgs~\cite{Bezrukov:2007ep} field, which is taken to behave as a spectator field. We treat the decay rate of the inflaton $\Gamma_{\phi}$  as a time- and space-dependent quantity because of its relation to the value of the Higgs field. Perturbations in the Higgs field lead to variations in $\Gamma_{\phi}$, similar to those discussed for example in Ref.~\cite{Dvali:2003em}. We use the perturbed Einstein equations introduced by Ref.~\cite{Dvali:2003em} to derive the temperature fluctuations induced by the spatial dependence of the Higgs field. While Ref.~\cite{Karam:2020skk} has a similar parameterization of Higgs effects on reheating, the amplitudes of temperature fluctuations we calculate in this work are notably larger for equivalent choices of parameters and we are able to show that large temperature fluctuations are even possible in the case of perturbative reheating without the full effects of Higgs blocking.\footnote{We note our calculations are broadly consistent with those in Ref.~\cite{Dvali:2003em}, which dynamically track the growth of density perturbations on superhorizon scales from the end of inflation. As the perturbation spectrum in Ref.~\cite{Karam:2020skk} is, alternatively, calculated using the $\delta N$ formalism at the time of inflaton decay (approximated to be instantaneous), a detailed comparison is beyond the scope of this work.} Furthermore, in our framework it is straightforward to extend our analysis to the case of non-perturbative preheating, for which we use a similar set of equations to calculate the density fluctuations induced by the spatial dependence of (Higgsed) SM gauge boson masses.
 
In Sec.~\ref{sec:Background} we present equations which describe the evolution of energy densities in different Hubble patches which are characterized by different inflaton decay rates in each patch. In Sec.~\ref{sec:PertEqns} we show the corresponding perturbed equations describing the evolution of adiabatic matter and metric perturbations after the end of inflation. After elaborating on the dynamics of the Higgs field in Sec.~\ref{sec:Higgs}, we explain our methods for calculating the comoving curvature perturbation in the cases of perturbative and resonant inflaton decays, as well as its connection to the temperature anisotropies observed in the CMB in  Secs.~\ref{subsec:PbPHiggs}, \ref{subsec:PbPendens} and \ref{sec:bardeen}, respectively. Our numerical results are presented in Sec.~\ref{sec:perturbative} for perturbative inflaton decays and  Sec.~\ref{sec:gauge} for the case of reheating via resonant decays of the inflaton into gauge bosons. We summarize our findings and offer our prospects for future work in Sec.~\ref{sec:discussion}.

To guide the reader to our main results: Our primary results for the case of (spatially-dependent) perturbative inflaton decay to SM particles can be found in Fig.~\ref{fig5}.  This figure shows constraints on the parameters (inflaton decay rate to SM fermions when masses are negligible, Yukawa coupling of SM decay products with the SM Higgs boson, and self-coupling of the Higgs during inflation) obtained by requiring that the amplitude of temperature fluctuations do not exceed CMB observations. For the case of resonant preheating to SM Higgsed gauge bosons, our main results can be seen in Fig.~\ref{fig:gaugeresults}; one can see that for all reasonable parameter choices the density fluctuations (shown in terms of the Bardeen potential) are too large.

\section{Reheat processes}
\label{sec:reheatProcesses}
\subsection{Unperturbed equations}
\label{sec:Background}

We work using a flat Friedmann-Lema\^{\i}tre-Robertson-Walker (FLRW) metric, described by the line element
\begin{eqnarray}
	ds^2 = dt^2 - a^2(t)\,d\boldsymbol{x}^2\,,
	\label{eq:FLRW}
\end{eqnarray}
where $a(t)$ is the scale factor, ${\bf x}$ are spatial coordinates, and $t$ is  cosmic time. We assume that towards the end of the inflationary period, the inflaton field begins to oscillate about the minimum of its potential, behaving as a massive scalar field with energy density ${\rho}_{\phi}$. We consider a collection of $n$ Hubble patches, each of which is characterized by a different inflaton decay rate $\Gamma^j_{\phi}$, for $j \in \{1, 2, \dots, n\}$. The perturbative decay of the inflaton field into radiation at the rate $\Gamma^j_{\phi}$ in the $j$-th Hubble patch is described by the coupled first-order equations
\begin{eqnarray}
	\frac{d{\rho}^j_{\phi}}{dN^j}=-3{\rho}^j_{\phi}-\frac{{\Gamma}^j_{\phi}}{H^j}  {\rho}^j_{\phi}\,,\label{eq:Boltzmann_mN} \\
	\frac{d{\rho^j}_r}{dN^j}=-4{\rho}^j_r+\frac{{\Gamma}^j_{\phi}}{H^j} {\rho}^j_{\phi}\,, 
	\label{eq:Boltzmann_rN}
\end{eqnarray}
where ${\rho}^j_{\phi}$, ${\rho}^j_r$ are the respective energy densities of the inflaton and of radiation, both in the same patch. The independent variable $N^j$ in Eqs.~\eqref{eq:Boltzmann_mN}-\eqref{eq:Boltzmann_rN} is the number of $e$-folds since the beginning of the reheating stage in patch $j$ at time $t_i$, defined as
\begin{eqnarray}
	N^j = \int_{t_i}^t H^jdt'\,,
\end{eqnarray}
where the Hubble rate is $H^j = \dot a^j/a^j$ with $a^j$ the scale-factor. The Hubble rate is related to the total energy density at the patch we are considering through the Friedmann equation
\begin{eqnarray}
	\left(H^j\right)^2 = \frac{8\pi G}{3}\left(\rho^j_{\phi}+\rho^j_r\right)\,,
	\label{eq:Hubble}
\end{eqnarray}
where $G$ is Newton's constant.

At this point it is also useful to define the background energy densities, with respect to which perturbations are calculated in Sec.~\ref{sec:PertEqns}. These background quantities are not evaluated at a particular Hubble patch like the ones mentioned above, but are in fact averaged over all Hubble patches (spatially independent) and only evolve with time. Their definitions are similar to those of Eqs.~\eqref{eq:Boltzmann_mN}-\eqref{eq:Boltzmann_rN} but include a decay rate $\bar{\Gamma}_{\phi}$, also averaged over all Hubble patches, as follows
\begin{eqnarray}
\frac{d{\bar{\rho}}_{\phi}}{dN}=-3\bar{\rho}_{\phi}-\frac{\bar{\Gamma}_{\phi}}{H}  \bar{\rho}_{\phi}\,,\label{eq:bar_Boltzmann_mN} \\
\frac{d\bar{\rho}_r}{dN}=-4\bar{\rho}_r+\frac{\bar{\Gamma}_{\phi}}{H} \bar{\rho}_{\phi}\,.
\label{eq:bar_Boltzmann_rN}
\end{eqnarray}
The Hubble parameter in this case is given by 
\begin{eqnarray}
H^2 = \frac{8\pi G}{3}\left(\bar{\rho}_{\phi}+\bar{\rho}_r\right)\,,
\label{eq:bar_Hubble}
\end{eqnarray}
and the number of $e$-folds is
\begin{eqnarray}
N = \int_{t_i}^t H dt'\,.
\label{eq:efolds}
\end{eqnarray}

\subsection{Perturbed equations} \label{sec:PertEqns}

In this Section we consider the evolution of the adiabatic matter and metric perturbations over the background quantities defined above. 

Working in the Newtonian gauge, we perturb the metric  of Eq.~\eqref{eq:FLRW} in the absence of  anisotropic pressure perturbations
\begin{eqnarray}
\label{eq:gravpot}
	ds^2 = (1+2\Phi)dt^2 - a^2(t)(1-2\Phi)d\boldsymbol{x}^2 \, ,
\end{eqnarray}	
where we introduced the gravitational potential perturbation  $\Phi$. We also write the perturbations in the energy density of the inflaton and  radiation components at a Hubble patch $j$ as
\begin{eqnarray}
	\rho^j_\phi = \bar{\rho}_\phi\,\left(1 + \delta^j_\phi\right), \qquad \rho^j_r = \bar{\rho}_r\,\left(1 + \delta^j_r \right)\,,
	\label{eq:perturbation_rho}
\end{eqnarray}
where $\delta^j_\phi, \delta^j_r \ll1$ are small perturbations over the averaged background quantities $\bar \rho_\phi$ and $\bar \rho_r$ respectively, at the particular patch. Despite the fact that, in principle, the perturbations themselves are dependent on the patch at which they are calculated, in the following we are going to refrain from using the $j$ superscript in our Equations. That is because, as we will explain in Sec.~\ref{subsec:PbPHiggs}, in our calculation we will only trace a characteristic value of the perturbations and not their actual distributions. This way we are later on able to constrain parameter space in a much more computationally efficient way. The actual perturbation distributions are calculated in our companion paper~\cite{Litsa:2020mvj} with the purpose of deriving the corresponding non-gaussianity of the temperature anisotropy spectrum.

The gravitational potential perturbation $\Phi$ couples to the energy density perturbations through
\begin{eqnarray}
	\frac{d\Phi}{dN} = -\Phi -\frac{4\pi G}{3H^2}\left(\bar{\rho}_{\phi} \delta_{\phi} + \bar{\rho}_r \delta_r\right)\,.
	\label{eq:delta_Hubble_N}
\end{eqnarray}
We now turn to the first order perturbations to the Boltzmann Eqs.~\eqref{eq:Boltzmann_mN}-\eqref{eq:Boltzmann_rN}, allowing also for fluctuations  $\delta \Gamma_{\phi}$ in the inflaton decay rate. We follow the method used in Ref.~\cite{Dvali:2003em} to assess the effects of such a perturbation. On super-horizon scales $k \ll aH$, the expressions describing perturbations in the matter and radiation energy densities read
\begin{eqnarray}
	\frac{d\delta_{\phi}}{dN} &=& 3\frac{d\Phi}{dN}-\frac{\bar{\Gamma}_{\phi}}{H}  (\delta_{\Gamma}+\Phi)\,, \label{eq:delta_Boltzmann_m_N}\\
	\frac{d\delta_r}{dN} &=& 4\frac{d\Phi}{dN} + \frac{\bar\rho_\phi}{\bar\rho_r}\frac{\bar{\Gamma}_{\phi}}{H} \left(\delta_{\Gamma} + \Phi +\delta_{\phi}-\delta_r\right)\,,\label{eq:delta_Boltzmann_r_N}
\end{eqnarray}
where $\delta_{\Gamma}\equiv \delta\Gamma_\phi/\bar\Gamma_\phi$ is the perturbation in the inflaton decay rate. The background Eqs.~\eqref{eq:Boltzmann_mN}-\eqref{eq:Boltzmann_rN} together with Eqs.~\eqref{eq:delta_Hubble_N}-\eqref{eq:delta_Boltzmann_r_N} form a set of five coupled differential equations.

In the standard scenario where $\delta_{\Gamma}=0$, super-horizon perturbations remain frozen until they re-enter the horizon (${d\Phi}/{dN} =0$). In this work, where there is modulated reheating, on the other hand, $\delta_{\Gamma} \neq 0$ leading to ${d\Phi}/{dN} \neq 0$; thus superhorizon perturbations evolve with time. In fact, the decay rate becomes time and space-dependent once the effects of \textit{Higgs modulation} and \textit{Higgs blocking} (as defined in Sec.~\ref{sec:Introduction} and discussed in FSSV) are taken into account during the reheating process. An alternative way of understanding the super-horizon evolution of the potential perturbation is to consider the isocurvature perturbations temporarily produced by the inhomogeneous transfer of energy between the inflaton and the radiation bath across different Hubble patches. We elaborate more on this topic in Sec.~\ref{sec:bardeen}.

\subsection{Higgs field dynamics}\label{sec:Higgs}

Having defined a framework for inhomogeneous reheating in previous sections, we now consider the dynamics of the Higgs boson during and after inflation, which determine the evolution of density perturbations produced during Higgs-modulated reheating. In this section, we summarize the aspects of the Higgs dynamics most relevant for Higgs-modulated reheating, while a more comprehensive discussion can be found in FSSV. We assume the SM Higgs is minimally coupled to gravity and is a light spectator field during inflation. 

We take the background value of the Higgs doublet and its potential to be
\begin{eqnarray}
\Phi &=& \frac{1}{\sqrt{2}}\left(\bfrac{0}{h}\right),\\
V_H(h) &=& \frac{\lambda}{4}\left(\Phi^\dag\Phi - \frac{\nu^2}{2}\right)^2 \approx \frac{\lambda}{4}\,h^4, \label{eq:higgs_potential}
\end{eqnarray}    
where $\nu = 246\,$GeV and $\lambda$ is the quartic self-coupling, which is taken to be positive during inflation. Here $h$ is a real scalar field.

\begin{figure*}[h!]
	\includegraphics[width=1.0\linewidth]{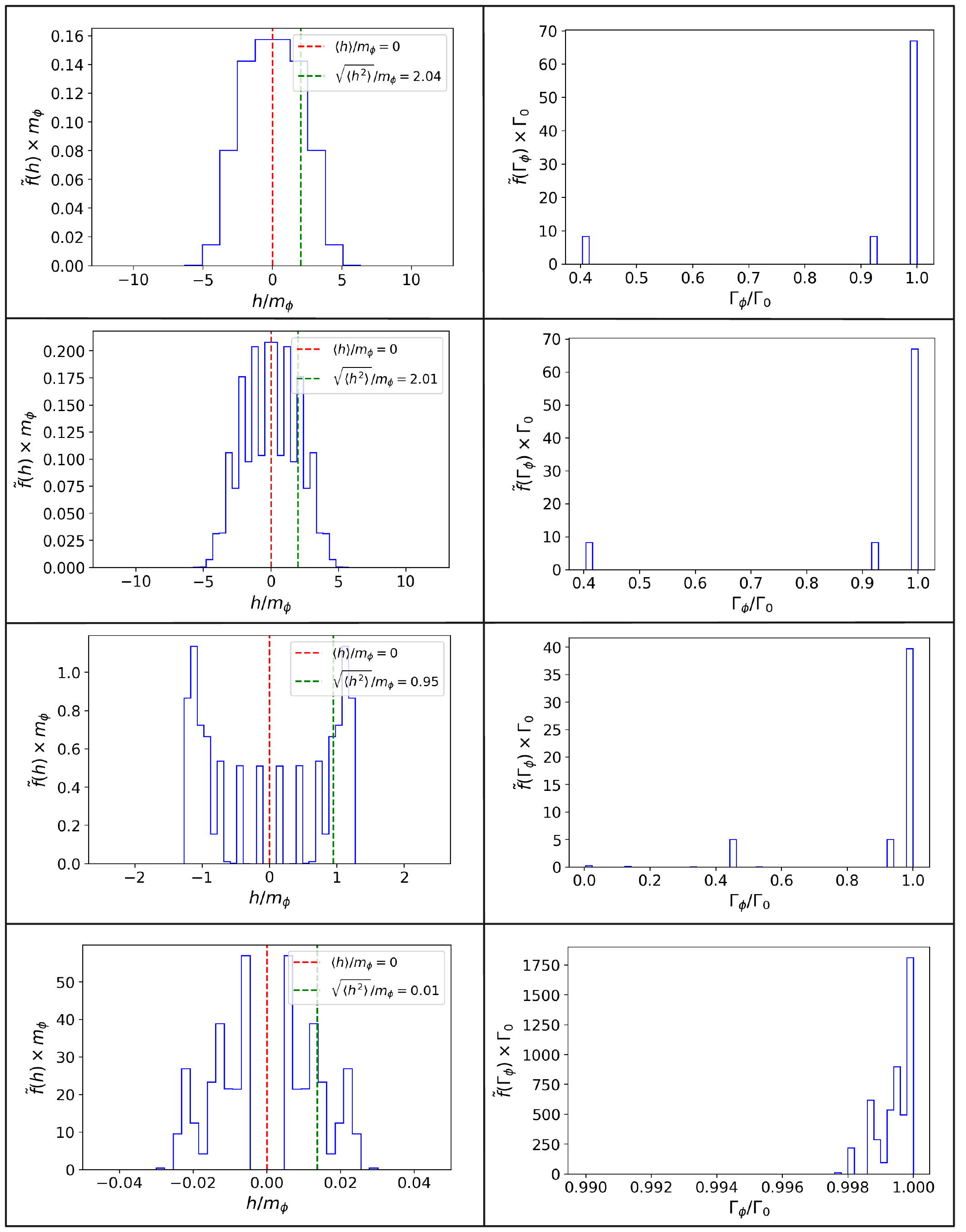} 
	\setlength{\abovecaptionskip}{-11pt}
	\caption{Probability Density Function of the Higgs (left panel) and, for the case of perturbative inflaton decay, the inflaton decay rate $\Gamma_{\phi}$ (right panel) at $N = 0$, $1$, $2.5$ and $6.5$ $e$-folds after the end of inflation from top to bottom. The dashed red and green lines correspond to the mean and root-mean-squared (standard deviation) values of the Higgs PDFs respectively. Both the Higgs (left panel) and the decay rate (right panel) values are binned. This figure is for $H_I = m_{\phi}$, $y = 1$, $\Gamma_0/m_{\phi} = 0.1$ and $\lambda_I = 10^{-3}$. Note that the scale of the x-axis changes for different values of N (top to bottom panels). One can see that the standard deviation value of the Higgs PDF decreases with increasing N, while the decay rates approach the unblocked value $\Gamma_0$ as blocking is lifted in more and more Hubble patches.} 
	\label{fig:PDFhG}
\end{figure*}

During inflation, the Higgs field initially rolls classically down its potential and soon reaches a regime dominated by quantum fluctuations. The result is that the super-horizon modes of the Higgs follow a random walk during the final stages of inflation. After a sufficient number of $e$-folds, the Probability Density Function (PDF)\footnote{Note that $\tilde{f}_{\rm eq}(h)$ has units of $[mass]^{-1}$, such that the Cumulative Distribution Function CDF $\equiv \int \tilde{f}_{\rm eq}(h) dh$ is dimensionless and equals unity when integrated over the entire domain $-\infty<h<\infty$. For example, see the axes of Fig.~\ref{fig:PDFhG}.} describing the Higgs field at the end of inflation is~\cite{Starobinsky:1994bd}
\begin{eqnarray}
\tilde{f}_{\rm eq}(h)=\left(\frac{32\pi^2\lambda_I}{3H_I^4}\right)^{1/4}\!\frac{1}{\Gamma (1/4)}\exp \!\left(\!-\frac{2\pi^2 \lambda_I h^4}{3H_I^4} \!\right).
\label{eqII8}
\end{eqnarray}
In the previous equation, $\lambda_I$ and $H_I$ are the Higgs quartic self-coupling and the Hubble scale at the end of inflation, respectively. Furthermore, the Gamma function has the value $\Gamma (1/4) \simeq 3.625$. 

Due to the stochastic dynamics mentioned above, each Hubble patch at the end of inflation has a different effective Higgs VEV. The probability to find a particular Higgs VEV in a given Hubble patch at the end of inflation is given by the equilibrium PDF of Eq.~\eqref{eqII8}. The reheating dynamics within each Hubble patch (after inflation has ended) are completely deterministic, once we specify the initial Higgs VEV value sampled from Eq.~\eqref{eqII8}. The dynamics of the Higgs field's VEV after inflation has ended in a Hubble patch $j$ reads
\begin{eqnarray}
\frac{d^2 h^j}{d\left(N^j\right)^2} + \left(3 + \frac{1}{H^j}\frac{dH^j}{dN^j}\right)\frac{d h^j}{dN^j} + \frac{\lambda_I}{\left(H^j\right)^2}  \left(h^j\right)^3 =0\,,\,\,\,\,\,\,\,\,\,
\label{eq:Higgs_N}
\end{eqnarray}
where the derivative of the Hubble rate with respect to the number of $e$-folds is
\begin{eqnarray}
\frac{dH^j}{dN^j} = \frac{1}{\sqrt{H^j}}\frac{4\pi G}{3}\left(\frac{d\rho^j_{\phi}}{dN^j}+\frac{d\rho^j_r}{dN^j}\right)\,.
\label{eq:dhdN}
\end{eqnarray}
For the moment, we have neglected the backreaction of the SM gauge bosons on the Higgs field dynamics, which is considered in Sec.~\ref{sec:backreaction}. Eqs.~\eqref{eq:Higgs_N}-\eqref{eq:dhdN} describe the damped oscillations which the Higgs experiences during the reheating period~\cite{Freese:2017ace}.  

Using the distribution of Higgs values at any point in time, we can define a characteristic value of the Higgs. The mean value of the Higgs VEV $\langle{h}\rangle$ across the observable Universe is zero, since for every patch with a positive value $h_j>0$ there will be another patch where the Higgs VEV has the value $-h_j$. We therefore take the characteristic value of the Higgs within a typical patch to be the standard deviation 
\begin{equation}
	\label{eq:higgsvev}
	\tilde{h} = \sqrt{\langle{h^2}\rangle} \equiv \left(\int h^2 \tilde{f}(h) dh\right)^{1/2} .
\end{equation}
Here $\tilde{h}$ is not the second moment of the equilibrium Higgs distribution in Eq.~\eqref{eqII8}. Instead, it is a time dependent quantity and must be evaluated from the actual Higgs distribution at every point in time. Specifically we will compute $\tilde{h}(N)$ as a function of the number of $e$-folds $N$ after inflation. Henceforth we will use the language \enquote{characteristic Higgs VEV} for this quantity $\tilde{h}$ (although not actually a VEV itself, but rather a typical value within a given patch).

At this point we should comment on the validity of Eq.~\eqref{eqII8} as the Higgs distribution at the end of inflation. $\tilde{f}_{\rm eq}$ represents the equilibrium PDF of any light spectator field present during inflation with a quartic self-interaction term dominating its potential at large field values (in our case the Higgs). The equilibrium PDF has been derived under two main assumptions. First, the calculation assumes a de-Sitter spacetime during inflation even though the small scale variation of the observed CMB power spectrum suggests a slight deviation from the pure de-Sitter limit. We also assume there is a sufficient number of $e$-folds during inflation $N_{\rm equil} \sim {\cal O}(\lambda^{-1/2})$ for the Higgs PDF to evolve towards equilibrium. 

While Eq.~\eqref{eqII8} is sufficient to both demonstrate the important effects of Higgs dynamics on reheating and derive interesting constraints, a more thorough stochastic analysis of the Higgs dynamics could provide further insight and even better (possibly model-dependent) constraints. Recent analysis of stochastic dynamics of light spectator fields has uncovered interesting results; for example Ref.~\cite{Hardwick:2017fjo} showed that light spectator fields can acquire larger field displacements during inflation when accounting for deviations from the de-Sitter approximation. If applied to our calculations, the parameter space of reheating models could be more tightly constrained than what is shown in Figs.~\ref{fig5} and~\ref{fig:gaugeresults}. 

Furthermore, the existence of four degrees of freedom in the Higgs field (ignoring gauge bosons for a moment), means that the Higgs random walk will be four-dimensional, leading to larger VEV's than its one-dimensional counterpart. The existence of gauge fields complicate the actual calculation, but as shown in Ref.~\cite{Adshead:2020ijf}, the end result for the Higgs VEV is closer to that of a four-dimensional random walk than the one-dimensional system leading to Eq.~\eqref{eqII8}. Taking a conservative viewpoint, we use Eq.~\eqref{eqII8} as the basis of our calculations, keeping in mind that  incorporating a more realistic PDF for the Higgs field will  result in even tighter constraints. 

One other concern we should address is the stability of the SM Higgs potential at large field values. The value of the Higgs quartic self-coupling $\lambda \equiv \lambda(\mu)$ depends on the renormalization parameter $\mu$ and can become negative at a high scale $\mu_{\rm inst} \sim 10^{11}$GeV due to its renormalization group (RG) evolution, possibly leading to an instability in the potential. The random walk of the Higgs field during inflation could thus send the Higgs into an anti-de Sitter minimum~\cite{EliasMiro:2011aa, Herranen:2014cua, Espinosa:2015qea, Herranen:2015ima}. This instability is sensitive to the value of the top quark mass $m_t$ which, for its best fit values leads to a negative Higgs potential above a (gauge-dependent) instability scale $\Lambda_{\rm inst} \approx 10^{11}\,$GeV. For simplicity, we assume that any possible instability in the Higgs potential is cured either by new physics decoupled from the inflation scale or by displacing the value of $m_t$ below its best fit value, within its significant experimental and theoretical uncertainties~\cite{Bednyakov:2015sca}.

\subsection{Perturbative decay: patch-by-patch Higgs evolution}
\label{subsec:PbPHiggs}

In the previous sections we defined both the background and the perturbation equations which govern the growth of inhomogeneities in the case of an inflaton decay rate which varies between different Hubble patches. Furthermore, we showed how the VEV of the Higgs field can differ from patch to patch due to its random walk during inflation. We now move on to explain the dependence of the inflaton decay rate on the Higgs VEV, which causes variation in reheating between Hubble patches. In this subsection, we consider the case of perturbative decay of the inflaton. In Sec.~\ref{subsec:PbPendens} we will turn to non-perturbative decay.

For simplicity, in computing the decay rate, we assume a Yukawa-type coupling between the inflaton and the fermion to which it decays. In this case, the decay rate of the inflaton at patch $j$ was shown in our previous work in FSSV to be
\begin{eqnarray}
	\Gamma^j_\phi = \Gamma_0\,\left[1 \!-\! \frac{4\left(m^j_f\right)^2}{m_{\phi}^2}\right]^{3/2}\Theta\left[m_\phi^2 \!-\! 4\left(m^j_f\right)^2\right]\;,
	\label{eq:decayrate}
\end{eqnarray}
where $m_\phi$ is the inflaton mass and $\Gamma_0$ is the inflaton decay rate in the absence of the Higgs modulation/blocking (the massless fermion limit). The Heaviside function, defined as $\Theta(x) = 1$ for $x \geq 0$ and $\Theta(x) = 0$ for $x < 0$, accounts for the phase-space blocking due to large effective fermion masses. Although our calculations specifically assume a Yukawa-type coupling between the inflaton and SM fermions, the results we present in the case of perturbative inflaton decay should remain basically the same for any final state particles that become massive due to the SM Higgs.  The details of the phase space of the decay products would vary for different inflaton-SM interactions or choice of final states but would not substantially change our results. 

The mass that the fermion acquires due to the Higgs mechanism in a patch with Higgs VEV $h^j$ is
\begin{eqnarray}
	m^j_f = \frac{y}{\sqrt{2}}\, h^j\, ,
\label{eq:fermionmass}
\end{eqnarray}
where $y$ is the Yukawa coupling for a given SM fermion. Although SM Yukawa couplings are technically scale dependent parameters, the RG evolution is typically insignificant between the electroweak and inflation scales for the minimal field content we consider in our analysis (i.e. the inflaton is the only new particle in addition to the SM). While we do consider the non-trivial RG evolution of other relevant SM parameters, in particular that of the Higgs quartic coupling $\lambda$ (as discussed in Secs.~\ref{sec:Higgs} and~\ref{sec:perturbative}), we assume the SM Yukawa couplings at the inflation scale are equivalent to the corresponding values at the electroweak scale. We thus present our results in terms of a generalized Yukawa coupling but make specific interpretations based on the hierarchy of Yukawa couplings relevant for the pattern of electroweak symmetry breaking in the SM.

Fig.~\ref{fig:PDFhG} presents the PDFs of the binned Higgs and inflaton decay rate values across all Hubble patches at $N = 0$, $1$, $2.5$ and $6.5$ $e$-folds after the end of inflation (from top to bottom), for $H_I = m_{\phi}$, $y = 1$, $\Gamma_0/m_{\phi} = 0.1$ and $\lambda_I = 10^{-3}$. Note that the scales of the (left panel's) horizontal axes have been chosen to vary in going from the top to the bottom panels of the figure, to accommodate the decreasing width of the Higgs PDF as a function of increasing $N$\footnote{Also see Fig.~\ref{fig:deltah}, where the blue solid line corresponds to $\delta h = \sqrt{\langle h^2\rangle}$.}. The decrease in the width of the Higgs PDF is due to the decreasing amplitude of the damped Higgs oscillations; according to Eq.~\eqref{eq:Higgs_N} the Higgs VEV becomes negligible within a few $e$-folds after the Higgs condensate begins to oscillate as a massive field. 
 
At the same time, the damping of Higgs oscillations causes the decay rates $\Gamma^j_{\phi}$ in the right panel to approach their unblocked value of $\Gamma_0 = \Gamma^j_{\phi}(h^j = 0)$. There are two effects that take place as the Higgs VEV $h^j$ in a Hubble patch decreases. First, the argument of the Heaviside function in Eq.~\eqref{eq:decayrate} becomes positive and Higgs blocking is lifted. After the lifting of Higgs blocking the decay rates are within the range $0< \Gamma^j_{\phi}/\Gamma_0 <1$. Also, Higgs modulation due to the second factor of Eq.~\eqref{eq:decayrate} (which we subsequently refer to as the {\it phase space factor}) approaches unity when the Higgs VEV has become much smaller than the mass of the inflaton. After both Higgs blocking and modulation are extinguished the decay rates are given by $\Gamma^j_{\phi}/\Gamma_0 = 1$. Because Higgs modulation/blocking are lifted at different times in different Hubble patches, in Fig.~\ref{fig:PDFhG} we can see some patches with $0< \Gamma^j_{\phi}/\Gamma_0 <1$, as well as others with $\Gamma^j_{\phi}/\Gamma_0 = 1$ at $N=6.5$ $e$-folds. In the following we will show that, even in the absence of Higgs blocking, Higgs modulation from the phase-space factor in the decay rate can lead to the production of large density perturbations. 

Furthermore, it is worth noting that the Higgs PDFs in the left panel of Fig.~\ref{fig:PDFhG} lose their original shape and become irregular after a sufficient amount of e-folds has passed. The reason is that Higgs oscillations proceed at different frequencies for different Higgs initial values $|h^j_{I}|$. However, since the oscillation frequency is the same for initial conditions with the same absolute value $|h^j_{I}|$, the Higgs PDFs remain symmetrical with respect to the vertical axis.

The ``average'' value of the decay rate $\bar \Gamma_\phi$ is obtained by using the characteristic value of the Higgs VEV $\tilde{h}$ from Eq.~\eqref{eq:higgsvev} and plugging into 
Eqs.~\eqref{eq:decayrate} and~\eqref{eq:fermionmass}. We find
\begin{eqnarray}
\begin{split}
\bar{\Gamma}_\phi &= \Gamma_0\,\left(1-\frac{2y^2\tilde{h}^2}{m_{\phi}^2}\right)^{3/2}\Theta\left(m_\phi^2 - 2y^2\tilde{h}^2\right)\\
& = \Gamma_0\,\left(1-\frac{2y^2\langle{h^2}\rangle}{m_{\phi}^2}\right)^{3/2}\Theta\left(m_\phi^2 - 2y^2\langle{h^2}\rangle\right)\;.
\label{eq:av_decayrate}
\end{split}
\end{eqnarray}
In principle, we could also define $\bar \Gamma_\phi$ in terms of $\langle h\rangle = 0$, which would give the constant value $\Gamma_0$. However, we are interested in calculating the density perturbations arising from the {\it relative} effects of Higgs modulation/blocking between different Hubble patches. Associated perturbations to the decay rate should thus be calculated with respect to the decay rate given by Eq.~\eqref{eq:av_decayrate} rather than the unblocked decay rate $\Gamma_0$, which would fail to capture any of the relevant modifications to the reheating dynamics.

The general expression for the perturbation of the decay rate relative to the average $\bar{\Gamma}_\phi$ can be calculated as
\begin{eqnarray}
\frac{\delta \Gamma_{\phi}}{\bar{\Gamma}_{\phi}} \!=\!
\begin{cases}
-\frac{6y^2\,h\delta h}{m_{\phi}^2}\left(1-\frac{2y^2\,h^2}{m_{\phi}^2}\right)^{-1}\!\!\!\!\!\!\!\!, & m_{\phi}^2> 4m_f^2\,\,\,\,\,\,\,\,\,\,\,\,\\
0\,, & m_{\phi}^2\leq 4m_f^2\,,\\
\end{cases}
\label{eq:deltaGamma_def}
\end{eqnarray}
where $\delta h$ is a variation of the Higgs VEV. We take $\delta h = \sqrt{\langle{h^2}\rangle}$ (corresponding to half of the distribution's width) and plug it into Eq.~\eqref{eq:deltaGamma_def}. The resulting characteristic perturbation is 
\begin{eqnarray}
\begin{split}
	\label{eq:decayrate_del}
	\delta_\Gamma &= \frac{\delta \Gamma_{\phi}}{\bar{\Gamma}_{\phi}}\Bigg|^{h \rightarrow \tilde{h}}_{\delta h \rightarrow \sqrt{\langle h^2\rangle}}= \\
	&= \begin{cases}
		-\frac{6y^2\,\hh}{m_{\phi}^2}\left(1-\frac{2y^2\,\hh}{m_{\phi}^2}\right)^{-1}\!\!\!, & m_{\phi}^2> 4m_f^2\,\\
		0\,, & m_{\phi}^2\leq 4m_f^2\,.\\
	\end{cases}
\end{split}
\end{eqnarray}

Our calculations to determine the decay rate perturbation $\delta_{\Gamma}(N)$ at every time-slice (labeled by the number of $e$-folds $N$) after the end of inflation proceed as follows. We will treat a large number $n \sim 10^4$ of Hubble patches, solving the equations independently for each patch. To obtain initial values for the Higgs field in multiple patches, we begin by drawing a sample of $n$ values of the Higgs field from the Higgs PDF in Eq.~\eqref{eqII8}, obtaining a collection of initial conditions $h^j_{I}$ with $j \in \{1, ... , n\}$.  Each value of the Higgs field corresponds to a different Hubble patch. Each value of $h^j_{I}$ is used as the initial condition for solving the system of four coupled differential Eqs.~\eqref{eq:Higgs_N}-\eqref{eq:dhdN},~\eqref{eq:Boltzmann_mN}-\eqref{eq:Boltzmann_rN} describing the Higgs and energy density evolution respectively. This results in a collection of $n$ solutions for the Higgs field $h_j= h_j(N)$, from which we calculate the value of $\sqrt{\hh}(N)$ that characterizes the Higgs distribution at every time slice.

Given $\tilde{h}(N) = \sqrt{\hh}(N)$, we determine $\bar{\Gamma}_{\phi}(N)$ and $\delta_{\Gamma}(N)$\footnote{An alternative way of determining these two functions would be to separately calculate the time-evolution of the decay rates $\Gamma^j_{\phi}$ ($j \in \{1, ... , n\}$) in $n$ Hubble patches. We could, then, define the average background decay rate as $\bar{\Gamma}_{\phi}(N) = \langle\Gamma_{\phi}\rangle$ and the decay rate perturbation as $\delta_{\Gamma} = \sqrt{\langle\Gamma^2_{\phi}\rangle}/\bar{\Gamma}_{\phi}$. A similar method is used in the case of resonant inflaton decay, as discussed in Sec.~\ref{subsec:PbPendens}. In the present section we choose to make the dependence on the distribution of Higgs VEVs manifest in order to highlight the role of the Higgs field in the perturbation dynamics. We have, however, verified that the two methods give results which agree within $\mathcal{O}(10 \%)$.} from respective Eqs.~\eqref{eq:av_decayrate} and \eqref{eq:decayrate_del} at every time slice (labeled by the number of $e$-folds $N$). To obtain the resultant density fluctuations, we then plug these functions $\bar{\Gamma}_{\phi}(N)$ and $\delta_{\Gamma}(N)$ into the background Eqs.~\eqref{eq:bar_Boltzmann_mN}-\eqref{eq:bar_Boltzmann_rN} and the perturbation Eqs.~\eqref{eq:delta_Hubble_N}-\eqref{eq:delta_Boltzmann_r_N}. Finally, we solve the system of five coupled differential Eqs.~\eqref{eq:bar_Boltzmann_mN}-\eqref{eq:bar_Boltzmann_rN}, \eqref{eq:delta_Hubble_N}-\eqref{eq:delta_Boltzmann_r_N} for the evolution of the gravitational potential $\Phi(N)$.  
 
To reiterate, in this section we introduced a method in which different Higgs VEVs in different Hubble patches  evolve separately and are averaged  at every time-slice in order for us to define the averaged background quantities and the corresponding perturbations. The benefit of this method is that it allows us to directly use Eqs.~\eqref{eq:decayrate}-\eqref{eq:decayrate_del} in conjunction with  Eqs.~\eqref{eq:delta_Boltzmann_m_N}-\eqref{eq:delta_Boltzmann_r_N}, to compute the density perturbations in matter (inflaton) and radiation during reheating. In other words, we are able to determine the time evolution of the inflaton decay rate and its perturbation, both of which are necessary components of the perturbed Einstein equations introduced in Ref.~\cite{Dvali:2003em}.   Results of applying this method to the case of perturbative inflaton decay will be presented in Sec.~\ref{sec:perturbative} below.

\subsection{Resonant decay: patch-by-patch energy densities}
\label{subsec:PbPendens}

The approach used in Sec.~\ref{subsec:PbPHiggs} to calculate the gravitational perturbation $\Phi$ is not applicable in cases where the inflaton decays non-perturbatively through parametric or tachyonic resonance. The reason is that in these cases we cannot define the decay rate ${\Gamma}_{\phi}$ as we did in the case of perturbative decay. Several models of parametric resonance have been studied since early works on the subject~\cite{Kofman:1997yn, Greene:1997fu}. Recent models include tachyonic preheating in $\alpha$-attractors~\cite{Krajewski:2018moi, Iarygina:2020dwe, Iarygina:2018kee}, Higgs inflation and related models~\cite{Sfakianakis:2018lzf, Ema:2016dny, vandeVis:2020qcp, Nguyen:2019kbm, DeCross:2016cbs}, as well as the formation of structures (such as oscillons) during preheating and their gravitational wave (GW) signatures~\cite{Lozanov:2019ylm}. To demonstrate the effects of Higgs modulation/blocking on preheating, we will consider the example of a model exhibiting tachyonic resonance~\cite{Felder:2000hj, ArmendarizPicon:2007iv}, specifically the case of an inflaton coupled to an abelian gauge field through a Chern-Simons term~\cite{Adshead:2015pva}. This paradigm is inspired by natural inflation~\cite{Freese:1990rb}, where the inflaton is an axion, possessing a shift-symmetry. Recent attention has focused on gauge field couplings to the inflaton potentially generating large scale magnetic fields~\cite{Adshead:2016iae} and a significant amount of GWs~\cite{Adshead:2018doq, Adshead:2019igv, Adshead:2019lbr}.

The effective Lagrangian for the system we consider is
\begin{eqnarray}
\nonumber
{\cal L} &=&
-{1\over 2} \partial_\mu \phi  \partial^\mu \phi -V(\phi)
\\
 &&-{1\over 4}  F_{\mu\nu}F^{\mu\nu} + {1\over 4 f} \phi F^{\mu\nu} \tilde F_{\mu\nu} +{M^2\over 2} A^\mu A_\mu
 \, ,
\end{eqnarray}
where $\phi$ is the inflaton and $f$ is proportional to the breaking scale of the associated $U(1)$ symmetry, expressed in units of the Planck mass $m_{\rm Pl}$. For the electromagnetic four-vector $A_\mu$, we have $F_{\mu\nu}= \partial_\mu A_\nu-\partial_\nu A_\mu$ and $\tilde F_{\mu\nu} = \epsilon_{\mu\nu\beta\gamma} F^{\beta\gamma}$ with the totally antisymmetric tensor $\epsilon_{\mu\nu\beta\gamma}$. We did not explicitly introduce the Higgs field in the above Lagrangian, but its effects are included through the gauge field mass, which is determined by $M=g |h|/2 $, where $g$ is the gauge coupling. We use abelian gauge fields as a proxy for the full electroweak SM sector, as explained in detail in FSSV, and leave a full analysis of non-abelian effects (see e.g. Ref.~\cite{Adshead:2017xll}) for future work.
Note that we refer to $1/f$ as the Chern-Simons coupling which is typically also proportional to the $U(1)$ charge of the inflaton and the square of the gauge coupling.

\begin{figure}[t]
	\includegraphics[width=1.0\linewidth]{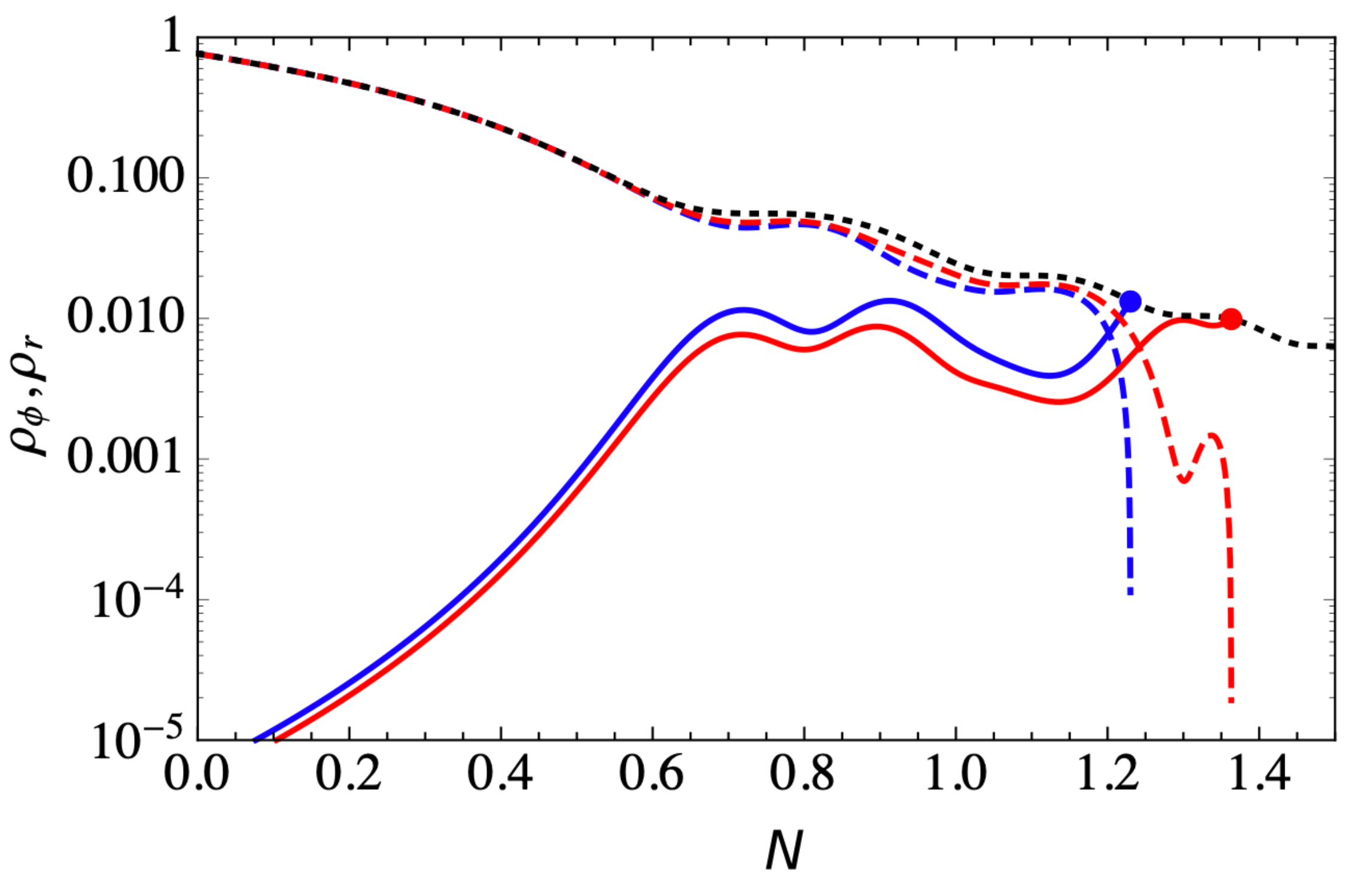} 
	\caption{For the case of resonant inflaton decay, the energy density budget as a function of $e$-folds $N$ after inflation for $f = 0.1\,m_{\rm Pl}$ and gauge field mass $M/m_\phi =0,1$ (blue and red respectively). We work in the linear fluctuation approximation, in which we neglect back-reaction effects. The solid curves correspond to the energy density in the produced gauge fields (radiation) $\rho_r$.  The black-dotted curve shows how the energy density of the inflaton would evolve without any transfer of energy to radiation from resonant particle production, $\rho_{\rm bg}$. The colored dashed lines show our approximation of the true inflaton energy density $\rho_\phi$ when we take into account the energy loss due to gauge field production. We approximate this as $\rho_\phi \equiv \rho_{\rm bg} - \rho_r$. The blue and red dots show our estimation for the time of complete preheating.} 
	\label{rhoparametric}
\end{figure}

For the calculation of the gravitational perturbations in the case of resonant decay we follow a slightly different method, focusing on the the transfer of energy from the inflaton to the gauge fields. We numerically solve the linearized equation of motion for the gauge field modes $A^\pm_k$, 
\begin{eqnarray}
	\ddot A_k^\pm + H \dot A^\pm_k + \left ({k^2\over a^2} \mp {k\over a} {\dot \phi\over H} + M^2\right )A^\pm_k=0 \, ,
	\label{eq:Ak}
\end{eqnarray}
where the superscript $\pm$ denotes the two helicities. Here $A_k = \chi_k^{1/2} a^{1/2}$ where $\chi_k$ is the Fourier transform of the transverse component of the gauge field. As discussed in our previous work FSSV, we consider the point where the energy density of the linearized fluctuations equals the energy density of the unperturbed inflaton background to be indicative of complete preheating. Although this approximation does not account for the back-reaction on gauge boson production, lattice simulations have shown that tachyonic resonance of this form can efficiently preheat the Universe~\cite{Adshead:2016iae}. 
 
Since tachyonic resonance is strong enough for the parameters chosen to completely preheat the Universe within ${\cal O}(1)$ $e$-folds, we neglect the evolution of the Higgs condensate, taking it to be fixed at the value it has at the end of inflation in each Hubble patch. By starting with a distribution of  Higgs values among different Hubble patches we define a similar distribution of gauge field masses $M$. By computing the energy density in radiation (gauge fields) and matter (inflaton condensate) in each patch (see Fig.~\ref{rhoparametric}), we define the averaged values $\bar \rho_\phi = \langle \rho_\phi \rangle$ and $\bar \rho_r= \langle \rho_r \rangle$  over all patches and the corresponding fluctuations $\delta\rho_{\phi} = \sqrt{\langle\rho^2_{\phi}\rangle}$ and $\delta\rho_{r} =\sqrt{\langle\rho^2_r\rangle}$ at each time-slice. Finally, the inflaton and radiation perturbations are given by the definitions  $\delta_{\phi} = \delta\rho_{\phi}/\bar{\rho}_{\phi}$ and $\delta_r = \delta\rho_{r}/\bar{\rho}_{r}$. Having calculated the energy density perturbation functions for the inflaton and radiation, we insert them into Eq.~\eqref{eq:delta_Hubble_N} to obtain the gravitational potential perturbation $\Phi$. The results of this method are shown in Sec.~\ref{sec:gauge}.

\subsection{Calculation of temperature anisotropies}
\label{sec:bardeen}

In order to constrain the allowed parameter space for reheating, taking the effects of Higgs modulation/blocking into account, we must connect our results to the temperature inhomogeneities observed in the CMB. We expect those to directly depend on the gauge-invariant comoving curvature perturbation, $\mathcal{R}$. On superhorizon scales, $\mathcal{R}$ is equivalent to the Bardeen parameter $\zeta$, defined as\footnote{Note that we have changed the overall sign of the definition by Ref.~\cite{Dvali:2003em} so that it exactly corresponds to the comoving curvature perturbation $\mathcal{R}$. The sign of the definition does not affect our results whatsoever, since in the following sections we are only tracking a characteristic value of the Bardeen parameter, rather than a distribution of perturbations which would include both over-densities and under-densities corresponding to hot-spots and cold-spots of the CMB.} ~\cite{Bardeen:1980kt, Bardeen:1983qw, Kodama:1985bj, Mukhanov:1990me, Baumann:2018muz, Dvali:2003em}
\begin{eqnarray}
	\label{eq:DefineBardeen}
	\zeta \!\equiv\! -\Phi \!+\! \frac{\bar{\rho}_{\phi}\delta_{\phi}\!+\!\bar{\rho}_r\delta_r}{3\bar{\rho}_{\phi}+4\bar{\rho}_r}\;.
\end{eqnarray}
Here $\Phi$ is the gravitational potential of Eq.~\eqref{eq:gravpot}, while $\bar{\rho}_{\phi}$ ($\bar{\rho}_{r}$) and $\delta_{\phi}$ ($\delta_{r}$) are the energy density background and perturbation on the background for the inflaton (radiation) respectively. 

In most models of inflation, the curvature perturbations are adiabatic and, thus, constant on superhorizon scales. However, in the case of modulated reheating, the situation is different. The curvature perturbations grow with time on superhorizon scales due to the spatial dependence of the inflaton decay. In order to gain intuition for this superhorizon growth, one can construct an unusual type of temporary isocurvature perturbation for superhorizon scales during reheating---the relative isocurvature perturbations between the inflaton and the radiation bath---as different amounts of energy are transferred from the inflaton to radiation in different Hubble patches,
\begin{eqnarray}
	\mathcal{S} = -3H\left(\frac{\delta \rho_{\phi}}{\dot{\bar\rho}_{\phi}} - \frac{\delta \rho_r}{\dot{\bar\rho}_r}\right)\;.
	\label{eq:isocurvature}
\end{eqnarray}
The associated time evolution of the Bardeen parameter is given by~\cite{Malik:2002jb}
\begin{eqnarray}
	\frac{\dot{\zeta}}{H} = \frac{1}{3}\frac{\dot{\bar{\rho}}_r\dot{\bar\rho}_{\phi}}{(\dot{\bar\rho}_r +\dot{\bar\rho}_{\phi})^2}\mathcal{S}\;.
	\label{eq:dot_zeta}
\end{eqnarray}

\begin{figure}[t]
	\includegraphics[width=1.0\linewidth]{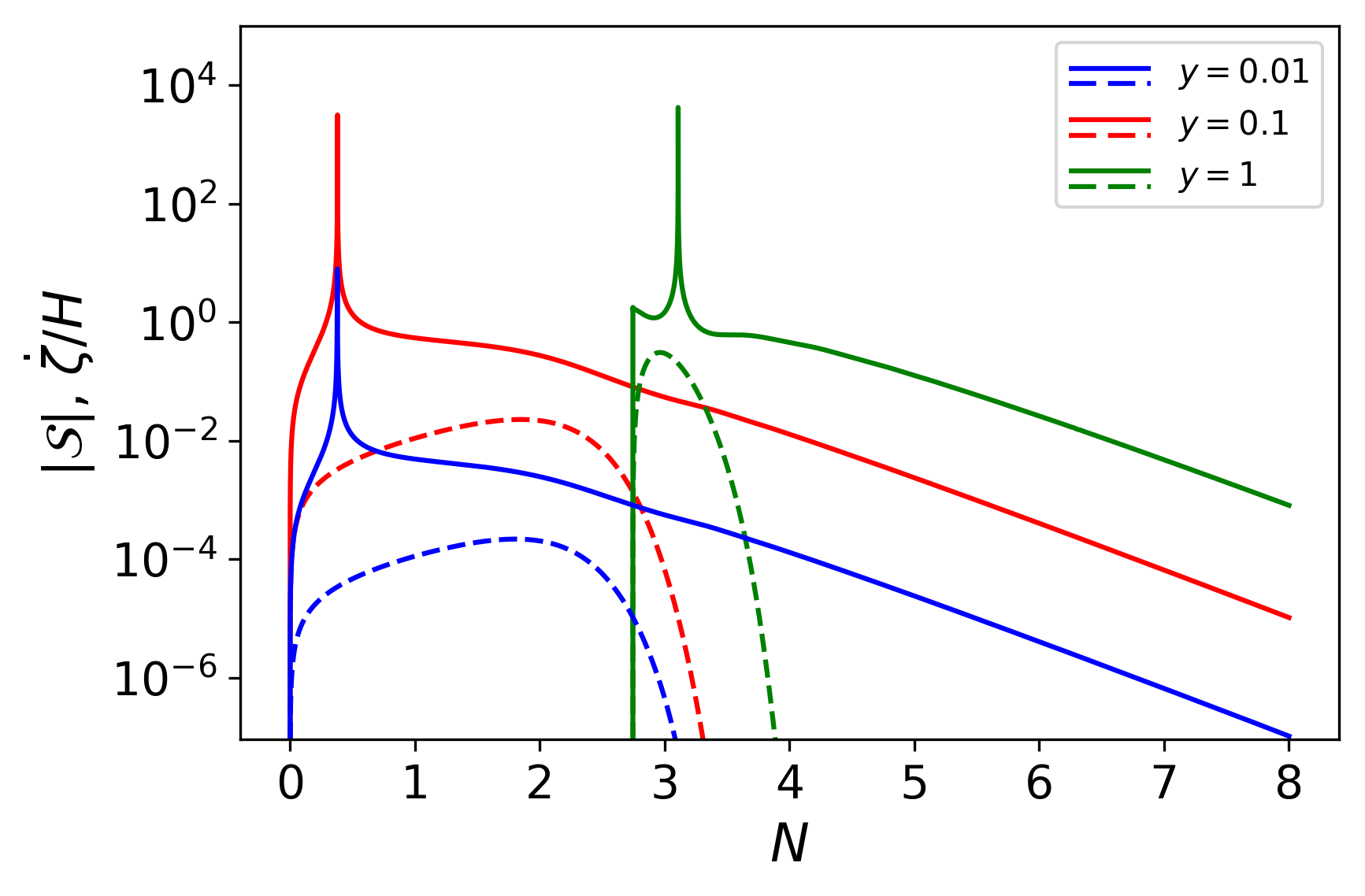} 
	\caption{The absolute value of the isocurvature perturbation, $|\mathcal{S}|$ (solid), and the time derivative of the Bardeen parameter re-scaled by the Hubble rate, $\dot{\zeta}/H$ (dashed), as a function of the number of $e$-folds after the end of inflation for the case of perturbative inflaton decay to fermions with Yukawa couplings $y = 0.01$ (blue), $y = 0.1$ (red) and $y = 1$ (green). We have used $\Gamma_0/m_{\phi} = 0.1$, $H_I = m_{\phi}$ and $\lambda_I = 10^{-3}$.}
	\label{fig:isocurvature}
\end{figure}

Note that these are not the usual isocurvature perturbations that lead to observables in the CMB (for example, the relative perturbations between radiation and matter that survive after the end of inflation). Instead, these are to be thought of as short-lived isocurvature perturbations that exist only during the reheating period. Furthermore, these isocurvature perturbations become rather ill-defined as the energy density of the inflaton vanishes in each Hubble patch. However, as mentioned above, it is informative to study their evolution during reheating to provide a better understanding for the way the Bardeen parameter grows on superhorizon scales. 

In Fig.~\ref{fig:isocurvature} we plot the time evolution of $\mathcal{S}$ (solid lines) and $\dot{\zeta}/H$ (dashed lines) calculated from respective Eqs.~\eqref{eq:isocurvature} and~\eqref{eq:dot_zeta} for the case of perturbative inflaton decay. Both $\mathcal{S}$ and $\dot{\zeta}/H$ are shown as functions of the number of $e$-folds after the end of inflation (the beginning of the inflaton oscillations). One can see the rapid initial increase of $\mathcal{S}$ (near the characteristic time of perturbative inflaton decay described in the following section) followed by  decay once energy is transferred to radiation. The time derivative of the Bardeen parameter re-scaled by the Hubble rate, $\dot{\zeta}/H$, rapidly decreases, approaching zero shortly after the end of inflation. As discussed further in Sec.~\ref{sec:Simplest_perturbative_reheating_case}, we have checked that taking the derivative of the Bardeen parameter in Eq.~\eqref{eq:DefineBardeen} gives the same result as the value of $\dot{\zeta}$ shown in Fig.~\ref{fig:isocurvature}. From either perspective, we can see that we are left with only adiabatic perturbations once every Hubble patch has completely reheated after inflation. Thus, the isocurvature fluctuations play no role in the calculation of CMB temperature anisotropies. While for simplicity we have discussed the temporary isocurvature perturbations associated with Higgs-modulated reheating in the case of perturbative inflaton decay, similar qualitative conclusions can be drawn in the case of reheating through resonant particle production.  

The exact relation between the temperature fluctuations and the gravitational potential or Bardeen parameter must also account for dynamics taking place at later times and, in particular, during the decoupling of CMB photons from the primordial plasma. On scales that remain outside of the horizon at the time of last scattering, the geodesics of CMB photons are altered by the distortions of spacetime due to matter perturbations in what is know as the \textit{Sachs-Wolfe} effect. Apart from the gravitational potential contributions, the full Sachs-Wolfe effect is calculated by taking into account perturbations intrinsic to the radiation plasma at the moment of photon decoupling. At linear order in the perturbations, the final result is expressed as~\cite{White:1997vi, Liddle:1993fq}
\begin{eqnarray}
\frac{\Delta T}{T} \!=\! \frac{1}{3}\Phi_f = \frac{1}{5}\zeta_f\;,
\label{eq:DT_T}
\end{eqnarray}
where $\Phi_f$ and $\zeta_f$ refer to the \textit{final} values of the gravitational potential and Bardeen parameter respectively at the time of CMB decoupling. We are only interested in the largest scales observable in the CMB, which re-enter the horizon well into the matter dominated epoch. Thus, we approximate that the amplitude of superhorizon perturbations calculated through the end of reheating are conserved through last scattering. 

\begin{figure*}
\begin{center}
	\includegraphics[width=1.0\linewidth]{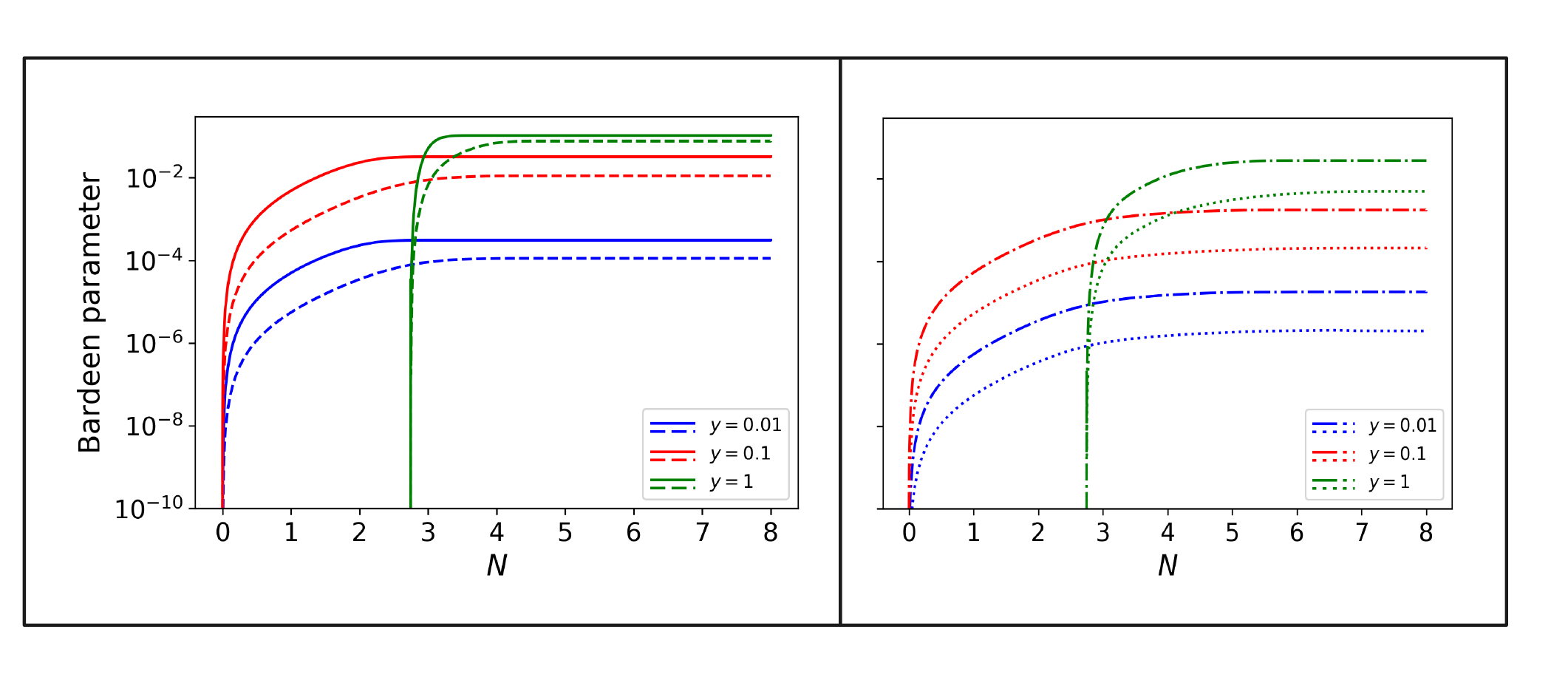} 
	\caption{For perturbative inflaton decay, the Bardeen parameter $\zeta$ is plotted as a function of  the number of $e$-folds after the end of inflation $N$ for three different values of the fermion Yukawa coupling: $y = 10^{-2}$ (blue), $y = 10^{-1}$ (red) and $y = 1$ (green). Different line styles correspond to different values of the unblocked decay rate: in the left panel $\Gamma_0 = 10^{-1}\, m_{\phi}$ (solid), $\Gamma_0 = 10^{-2}\, m_{\phi}$ (dashed), and in the right panel $\Gamma_0 = 10^{-3}\, m_{\phi}$ (dot-dashed) and $\Gamma_0 = 10^{-4}\, m_{\phi}$ (dotted). We have set the Hubble scale at the end of inflation at $H_I = m_{\phi}$ and the Higgs self-coupling at $\lambda_I = 10^{-3}$.} 
\label{fig1}
\end{center}
\end{figure*}

\section{Results for the Case of Perturbative Inflaton Decay}
\label{sec:perturbative}

In this section we discuss the temperature fluctuations produced by Higgs-modulated reheating in the case of perturbative inflaton decay, derived using the method described in Secs.~\ref{subsec:PbPHiggs} and \ref{sec:bardeen}. Again, we assume a decay rate given by Eq.~(\ref{eq:decayrate}), corresponding to a Yukawa-type coupling between the inflaton and the fermion to which it decays. In Sec.~\ref{sec:Simplest_perturbative_reheating_case} we start with the simplest scenario in which we make the following two approximations: gauge bosons produced by the resonant decay of the Higgs condensate do not back-react on the Higgs evolution, as well as the assumption that the frequency of oscillations by Higgs the condensate is much slower than the Hubble rate. The effects of backreaction on the Higgs evolution and assuming the opposite limit of rapid Higgs oscillations will be investigated in Secs.~\ref{sec:backreaction} and~\ref{sec:Assuming time-averaged Higgs oscillations} respectively. Finally, in Sec.~\ref{sec:comparison} we present constraints on a generalized parameter space for perturbative reheating by requiring that the amplitude of temperature fluctuations not exceed that observed in the CMB and then describe how our results could be extrapolated to models with a lower inflation scale in Sec~\ref{sec:lowerscale}.

Before proceeding, we will briefly comment on the value of Higgs self-coupling at the inflation scale $\lambda_I$. Assuming no new physics couples to SM Higgs, RG evolution of  $\lambda$ between the electroweak and the inflation scale will yield $\lambda_I \simeq 10^{-2}$ at the end of high scale inflation~\cite{Degrassi:2012ry}. In the following sections we will primarily use $\lambda_I = 10^{-3}$ since a smaller $\lambda_I$ causes the production of slightly larger adiabatic density perturbations and intensifies effects such as the backreaction of gauge bosons on the Higgs dynamics. After assuming $\lambda_I = 10^{-3}$ in order to more clearly demonstrate various aspects of density perturbations produced by Higgs modulation/blocking, we will use $\lambda_I = 10^{-2}$ as a benchmark when constraining the parameter space of reheating.

\subsection{Simplest perturbative reheating case}
\label{sec:Simplest_perturbative_reheating_case}

As mentioned above, we start with the simplest scenario in which we ignore the effect of gauge boson backreaction on the evolution of the Higgs condensate and assume the frequency of the Higgs oscillations is slower than the Hubble rate. In Fig.~\ref{fig1} we present the Bardeen parameters $\zeta$ as a function of the number of $e$-folds after the end of inflation. We have equated the Hubble scale at the end of inflation to the inflaton mass $H_I = m_{\phi}$ and used the value $\lambda_I = 10^{-3}$ for the Higgs self-coupling. 

As noted above, we see the growth of the Bardeen parameters coincide with that of the corresponding isocurvature perturbations in Fig.~\ref{fig:isocurvature}. More specifically, we can see the Bardeen parameter $\zeta$ of Fig.~\ref{fig1} stabilizes at the same time that its derivative $\dot{\zeta}/H$ in Fig.~\ref{fig:isocurvature} approaches zero. We have also checked that numerically differentiating the Bardeen parameters of Fig.~\ref{fig1} with respect to time yields the same values of $\dot{\zeta}$ as those shown in Fig.~\ref{fig:isocurvature}. Thus, we conclude that the growth of adiabatic density perturbations on superhorizon scales can indeed be described as the temporary generation of isocurvature modes during the inhomogeneous reheating process.

We find two main results, which are clearly demonstrated in the simplest case of perturbative reheating but also generally hold under more complicated assumptions discussed in subsequent sections. First, we find that the perturbations caused by Higgs modulation/blocking, $\Delta T/T|_{\text{H}} \sim \zeta/5$, are larger for larger values of  the Yukawa coupling $y$ and the decay rate $\Gamma_0$. In fact, they can exceed the temperature fluctuations observed in the CMB $\Delta T/T|_{\text{CMB}}\sim 10^{-5}$ by several orders of magnitude for certain parameter combinations. For example, we find $\Delta T/T|_{\text{H}} \gtrsim {\cal O}(10^{-4})$ for $\lambda_I = 10^{-3}$, $y \gtrsim \mathcal{O}(10^{-1})$ and $\Gamma_0/m_{\phi}\gtrsim {\cal O}(10^{-4})$.  An extensive examination of the full parameter space $(\Gamma_0, y, \lambda_I)$ and bounds from the CMB will be presented in Sec.~\ref{sec:comparison}.

Second, we see that Higgs blocking of the inflaton decay into fermions takes place only for $y\gtrsim 1$.  In Fig.~\ref{fig1}, we observe that the Bardeen parameter in the case of $y=1$ grows sharply only after $N\simeq 3$ $e$-folds, whereas for smaller values of $y$ the growth happens much sooner and more gradually. This trend is consistent with the results of FSSV, where we showed that large Yukawa couplings are needed to cause a significant delay of the reheating process.  In that paper we showed that the reheat temperature could be suppressed by up to an order of magnitude compared to the unblocked case. 

In Fig.~\ref{fig1} one can see that larger values of the inflaton decay rate lead to more rapid increase of the Bardeen parameter. This trend can be understood by comparing the timescale relevant for Higgs blocking to that of reheating, which can only be completed when $\Gamma_{\phi} \gtrsim H$.  For example, at large values of $y = 1$ and $\Gamma_0/m_{\phi} = 0.1$, the decay rate of the inflaton is already larger than the Hubble rate by the time Higgs blocking has been lifted. Hence, after blocking is no longer an obstacle, reheating takes place instantaneously and $\zeta$ increases sharply. On the other hand,  for $y=1$ and smaller decay rates (right panel, with $\Gamma_0/m_{\phi} =  10^{-3}, 10^{-4}$), the expansion rate is still larger than the decay rate when Higgs blocking is lifted. As a result, reheating requires more time and the increase of $\zeta$ happens more gradually.

For cases where $y\lesssim 0.1$, there is never any Higgs blocking; yet substantial density perturbations may still result simply due to Higgs modulation.  In other words, the Yukawa couplings are not sufficiently high to make the argument of the Heaviside function in Eq.~\eqref{eq:decayrate} negative, and thus do not block reheating (or block reheating only at an exponentially suppressed number of rare Hubble patches). However, whether or not Higgs blocking is ever significant, there is always a residual dependence of the phase space factor on $h$, which creates differences between decay rates at different Hubble patches. This \textit{Higgs modulation} can thus still lead to the production of large density perturbations. 

It might be possible that, due to the deviation from a pure de-Sitter spacetime during inflation, the equilibrium distribution of Higgs VEVs given by Eq.~\eqref{eqII8} is not a valid approximation and larger field values could be expected from the quantum fluctuations of the spectator Higgs field~\cite{Hardwick:2017fjo}. For larger Higgs VEVs across a significant number of Hubble patches, smaller Yukawa couplings  would be sufficient to produce the same results calculated under the assumption of a de-Sitter equilibrium distribution. More specifically, the curves shown in Fig.~\ref{fig1} for a given combination of $(\Gamma_0, y, \lambda_I)$ can be approximately re-interpreted as corresponding to the evolution of the Bardeen parameter with $(\Gamma_0, y', \lambda_I)$, where the re-scaled Yukawa coupling is given by $y' = (\tilde{h} / \tilde{h} ') y$ for the modified characteristic Higgs VEV $\tilde{h}'$. While a detailed analysis of the changes to our results for deviations of the Higgs PDF from Eq.~\eqref{eqII8} are beyond the scope of this work, the possibility of larger Higgs VEVs across a sufficient number Hubble patches would imply that the constraints on the parameter space of reheating in this paper should be considered conservative.

\subsection{Including the effects of backreaction on Perturbative Reheating}
\label{sec:backreaction}

The results discussed in the previous section were obtained by neglecting the effects of the gauge bosons which are produced resonantly from the decay of the Higgs condensate~\cite{Enqvist:2013kaa}. These gauge bosons can potentially back-react on the dynamics of the Higgs boson. In this section, we summarize the treatment of backreaction discussed in FSSV and show that the associated effects on the density fluctuations produced during Higgs-modulated reheating may be neglected.

{\it Gauge Boson Production:} The induced mass of the SM W-bosons $m_W =g|h|/2$ vanishes when the oscillating Higgs field crosses zero, leading to a substantial production of gauge boson particles~\cite{Starobinsky:1994bd, Degrassi:2012ry}. During the oscillations of the Higgs field, the transverse components of the SM gauge fields oscillate with a time-dependent frequency $\omega_k$ depending on the mode with wavenumber $k$, and with a corresponding occupation number $n_k$~\cite{Greene:1997fu, Enqvist:2013kaa}. The backreaction associated with the resonant production of the gauge bosons on the Higgs dynamics manifests as an effective mass squared term in the Higgs equation of motion, Eq.~\eqref{eq:Higgs_N}, given by
\begin{eqnarray}
	m_{h(W)}^2 = \frac{g^2}{4}\int \frac{\mathrm{d}^3{\bf k}}{(2\pi a)^3}\frac{n_k}{\omega_k}\,,
	\label{eq:gaugebosonmass}
\end{eqnarray}
where $g$ is the coupling constant appearing in the covariant derivative of the Higgs to the gauge field. The resonant production of the gauge bosons is governed by the quantity $q_W \equiv g^2/(4\lambda)$, see Appendix~\ref{appI} for further discussion. We ignore the non-Abelian self-interactions of the gauge fields~\cite{Enqvist:2014tta}, which may change the Higgs condensate decay time but would not drastically affect our overall results. 

{\it Backreaction:} The backreaction from gauge bosons takes effect when the effective mass squared of the gauge bosons $m_{h(W)}^2$ is of the same order as the effective mass squared $m_{h(\lambda)}^2$ of the Higgs field given by its self-coupling~\cite{Enqvist:2013kaa},
\begin{eqnarray}
	m_{h(W)}^2 \simeq  m_{h(\lambda)}^2 \equiv 3 \lambda h^2 \, .
	\label{eq:HiggsDecay}
\end{eqnarray}
When the condition in Eq.~\eqref{eq:HiggsDecay} is met, we assume the Higgs field instantaneously decays away and the inflaton decay rate---no longer blocked---becomes equal to $\Gamma_0$. The full decay rate of the inflaton, accounting for the effects of \textit{backreaction} (BR), is therefore~\cite{Freese:2017ace}
\begin{eqnarray}
\nonumber
	\Gamma_{\rm BR} &=& \bar{\Gamma}_{\phi}\,\Theta{\left(3\lambda_I h^2 - m_{h(W)}^2\right)} 
	\\ && + \Gamma_0\,\Theta{\left(m_{h(W)}^2 - 3\lambda_I h^2\right)}\;.
	\label{eqIII2}
\end{eqnarray}
The expression for $\Gamma_{\rm BR}$ above imposes that Higgs blocking only affects the dynamics during the time period when backreaction can be neglected, $m_{h(W)}^2 \leq 3\lambda_I h^2$. Following the analysis in FSSV, the time from the end of inflation when the Higgs condensate decays away due to backreaction effects reads
\begin{eqnarray}
	\label{eqIII3}
	t_{\rm dec} = t_{\rm osc} + \frac{1}{m_\phi}\,\left(\frac{\ln n_{k=0}}{\mu_0}\right)^2\;,
\end{eqnarray}
where the factor $\mu_0 = 0.185$ is obtained from a numerical fit to the exact solution for the time dependence of $n_k$~\cite{Freese:2017ace} and the occupation number at wave number $k=0$ is
\begin{eqnarray}
	n_{k=0} = \frac{9\pi\sqrt{2}}{\lambda_I}\,\left(\frac{2\pi^2}{q_W}\right)^{5/4}\;.
\end{eqnarray}
\begin{figure}[t]
 	\includegraphics[width=1\linewidth]{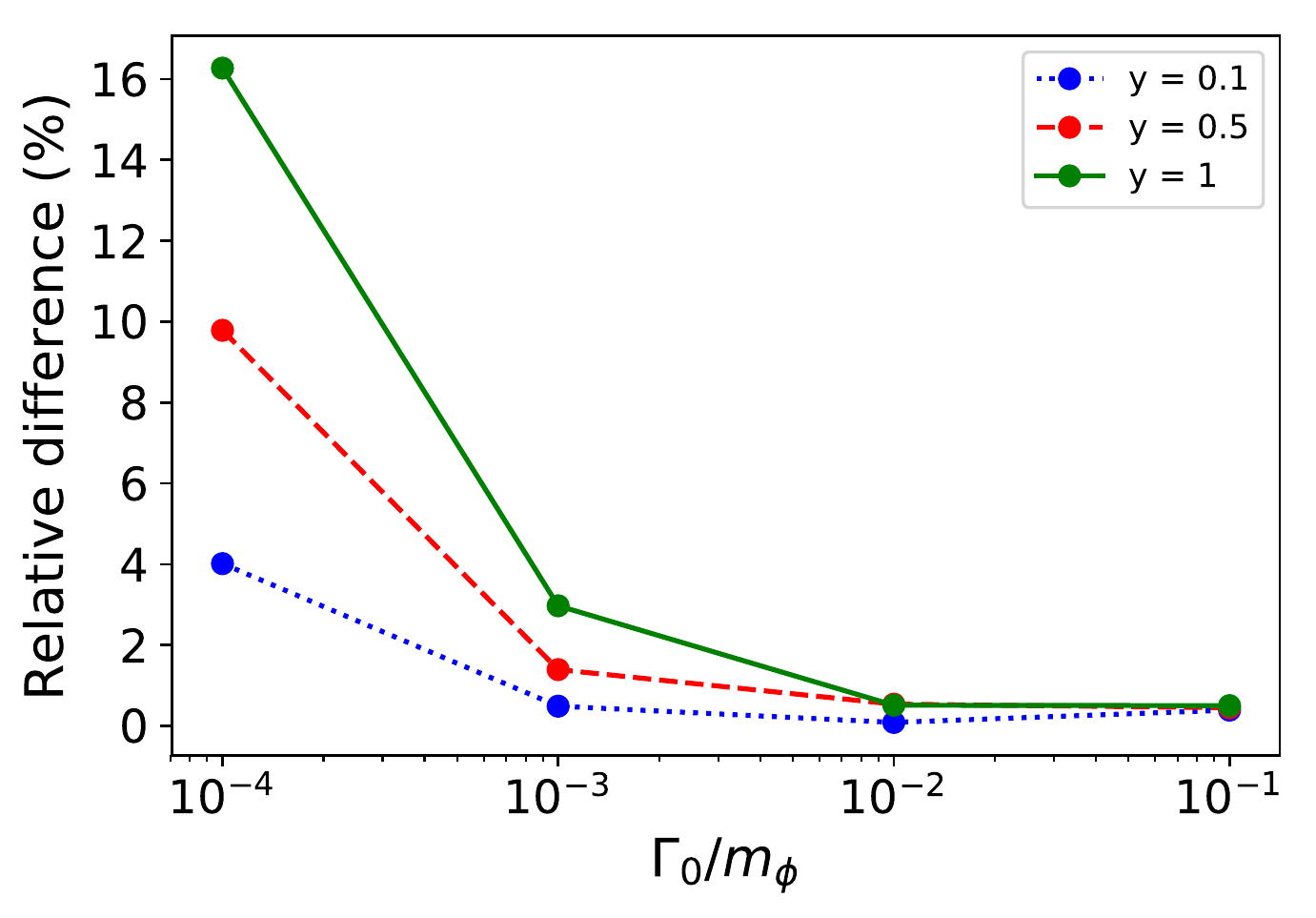} 
 	\caption{For the case of perturbative inflaton decay, the fractional difference between perturbations, $ (\zeta - \zeta_{\rm BR})/\zeta$, with ($\zeta_{\rm BR}$) and without ($\zeta$) including the effects of backreaction, as a function of the unblocked decay rate $\Gamma_0$, for three different Yukawa couplings $y=0.1$ (blue dotted), $y=0.5$ (red dashed), and $y=1$ (green solid). We assume $H_I=m_{\phi}$ and $\lambda_I=10^{-3}$.}
 	\label{fig2} 
\end{figure}

In the following we point out the conditions for which backreaction can impact the calculation of the density perturbations produced by Higgs-modulated reheating. First, the unblocked decay rate of the inflaton field $\Gamma_0$ must be low enough to ensure that the Higgs condensate will decay before reheating has been completed. In other words, since the \enquote{seed} of the perturbations from Higgs-modulated reheating is the difference in decay rates between different Hubble patches, only further modifications to the decay rates will cause the calculation of perturbations to change. The effects of backreaction can only modify the decay rate in a given Hubble patch if the Higgs field is given sufficient time to decay before reheating has been completed and the decay rate has become the same (equal to $\Gamma_0$) in every patch. More specifically, the relation between the unblocked inflaton decay rate $\Gamma_0$ and the Hubble parameter at the time of the Higgs condensate decay $H_{\rm dec}$, should be $\Gamma_0 \lesssim H_{\rm dec}$. 

The second condition for the effects of backreaction to be important requires that the Yukawa coupling $y$ is high enough, such that the decay rate of the inflaton does not become immediately equal to $\Gamma_0$ after inflation has ended. As mentioned earlier, if Higgs blocking never occurs the only source of the perturbations is the phase space factor causing slight differences in the decay rates of different patches. The smaller the Yukawa coupling is, the smaller the deviations from $\Gamma_0$ caused by the phase space factors are. As a result, for Yukawa couplings small enough such that the transition from $\bar{\Gamma}_{\phi}$ to $\Gamma_0$ is immediate once Higgs blocking is lifted, the effects of backreaction do not cause considerable modification of the decay rates. Therefore, the density perturbations calculated for small $y$ when accounting for the effects of backreaction, are not considerably different compared to those calculated in Sec.~\ref{sec:Simplest_perturbative_reheating_case} ignoring the effects of backreaction.

We demonstrate the conditions under which the effects of backreaction are relevant in Fig.~\ref{fig2}, where we plot the fractional difference between perturbations with and without backreaction $(\zeta - \zeta_{\rm BR})/\zeta$ as a function of the unblocked decay rate $\Gamma_0$, for the values of the Yukawa coupling $y = 0.1$ (blue dotted), $y = 0.5$ (red dashed), and $y = 1$ (green solid), while setting $\lambda_I = 10^{-3}$.  It is evident that changes to the size of density perturbations due to the effects of backreaction increase for larger Yukawa couplings and decrease for larger decay rates $\Gamma_0$, thus verifying our intuitive understanding described above. More specifically, deviations between the two methods approach zero for $\Gamma_0/m_{\phi}\gtrsim 10^{-2}$. Furthermore, the largest differences between the perturbations with and without backreaction, in the case of Yukawa couplings $y \gtrsim 1$, are well below an order of magnitude and require extremely small decay rates to become significant.

\begin{figure}[t]
	\includegraphics[width=1.0\linewidth]{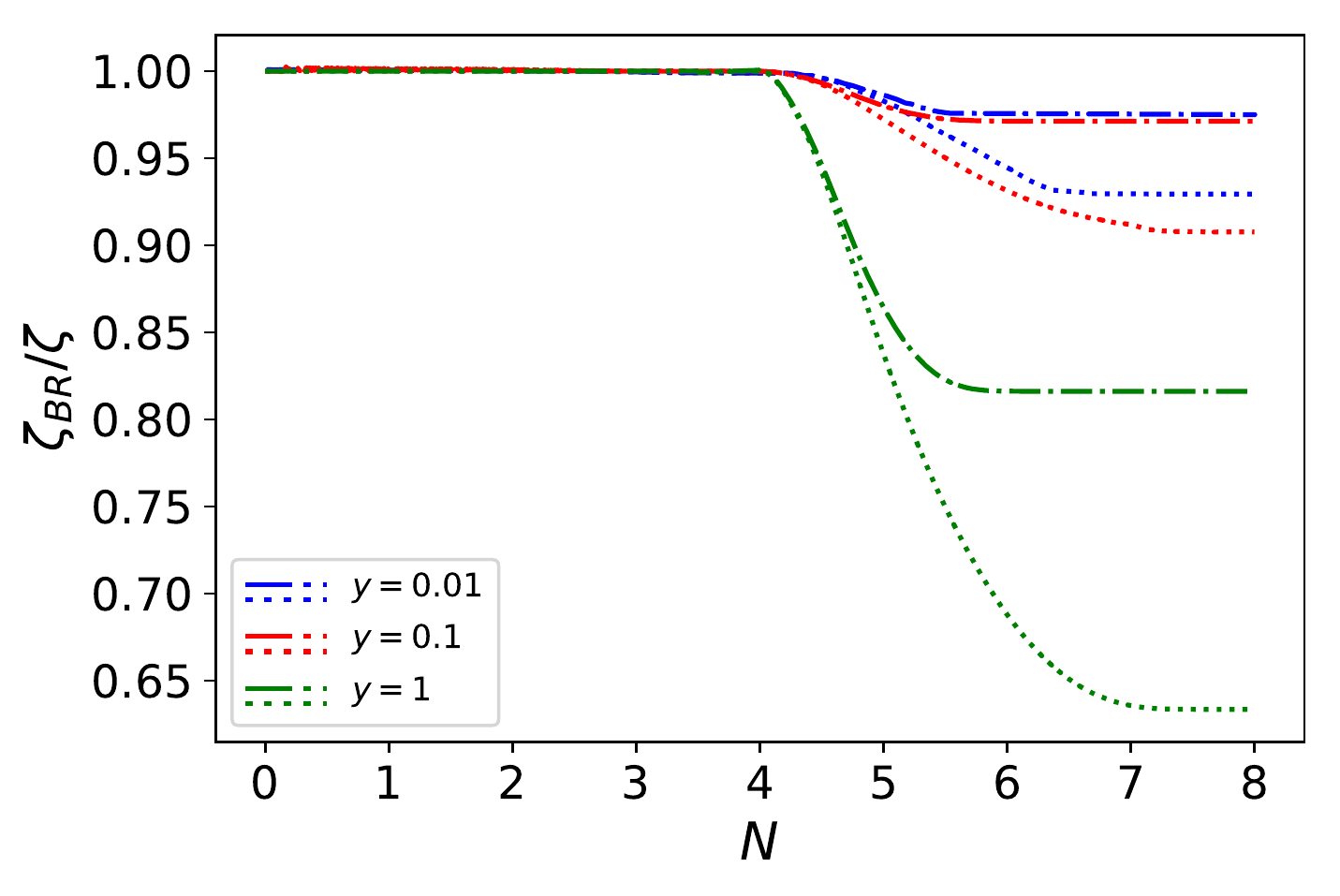} 
	\caption{For the case of perturbative inflaton decay, the ratio $\zeta_{BR}(N)/\zeta(N)$ of Bardeen parameter calculated with and without including the effects of backreaction as a function of the number of $e$-folds after the end of inflation for Yukawa couplings $y=0.01$ (blue), $0.1$ (red), and $1$ (green). Note the quantity $\zeta$ is plotted in the right panel of Fig.~\ref{fig1} without considering backreaction. The dash-dotted lines correspond to $\Gamma_0 = 10^{-3}\,m_{\phi}$, while the dotted lines to $\Gamma_0 = 10^{-4}\,m_{\phi}$. We assume $H_I=m_{\phi}$ and Higgs self-coupling $\lambda_I=10^{-3}$.}
	\label{fig3} 
\end{figure}

Fig.~\ref{fig3} compares the evolution of density perturbations with and without including the effects of backreaction. In order for the effects to be visible by eye, we only show examples with relatively low decay rates $\Gamma_0 = 10^{-3}\,m_{\phi}$ and $\Gamma_0 = 10^{-4}\,m_{\phi}$. We plot the ratio of the Bardeen parameters calculated when including the effects of backreaction to those in the right panel of Fig.~\ref{fig1} where backreaction was ignored. As expected from Fig.~\ref{fig2}, the largest modification of the perturbations occurs for the combination of the largest Yukawa coupling $y = 1$ with the smallest decay rate $\Gamma_0 = 10^{-4}\,m_{\phi}$. Since Yukawa couplings as large as $y = 1$ are only relevant for the case of the inflaton decaying into the SM through a coupling to the top quark and the associated perturbations are much larger than observed in that part of our reheating parameter space, we choose to ignore effects of backreaction in the following sections.\footnote{As we will see below there is a small portion of allowed parameter space with Yukawa couplings equal to $1$ for extremely small decay rates. However, considering the small size of this region in parameter space and the insignificance of the backreaction effect in any case, ignoring it remains a very good approximation.}

The conclusions about the effects of backreaction drawn so far have been based on calculations assuming a rather small value of the Higgs quartic coupling $\lambda_I = 10^{-3}$. However, the approximation we make by ignoring backreaction holds for larger self-couplings because the second term of Eq.~\eqref{eqIII3} contains a contribution proportional to the logarithm $\ln(\lambda_I)$~\cite{Freese:2017ace}. Thus, larger couplings $\lambda_I > 10^{-3}$ result in longer decay times for the Higgs condensate, $t_{\rm dec}$. The effects of backreaction for larger quartic couplings are, according to the first condition described above, only relevant for even smaller values of $\Gamma_0$ in order for the Higgs field to decay before reheating is completed. Choices of $\lambda_I > 10^{-3}$ further limit the portion of parameter space where the effects of backreaction are relevant and we are therefore justified in ignoring the effects of backreaction for cases with larger quartic couplings subsequently considered in this paper. 

\subsection{The case of rapid Higgs oscillations in Perturbative Reheating}
\label{sec:Assuming time-averaged Higgs oscillations}

After inflation has ended, the Higgs experiences damped oscillations. For simplicity, we calculate the Higgs evolution in each Hubble patch by considering two limiting cases where the oscillation period of the Higgs field $\THi$ is either much longer or much shorter than the Hubble time, $H^{-1}$. In the work presented so far, we have assumed the limit $\THi \gg H^{-1}$, which allows for the resolution of the zero-crossings in the oscillations of the Higgs that gradually cause Higgs modulation/blocking to be lifted~\cite{Freese:2017ace}. Under such an assumption, it is sufficient to directly sample values of the Higgs field from the distribution of Higgs VEVs computed at every time-slice, as described in Sec.~\ref{subsec:PbPHiggs}. 

\begin{figure}[t]
	\includegraphics[width=1.0\linewidth]{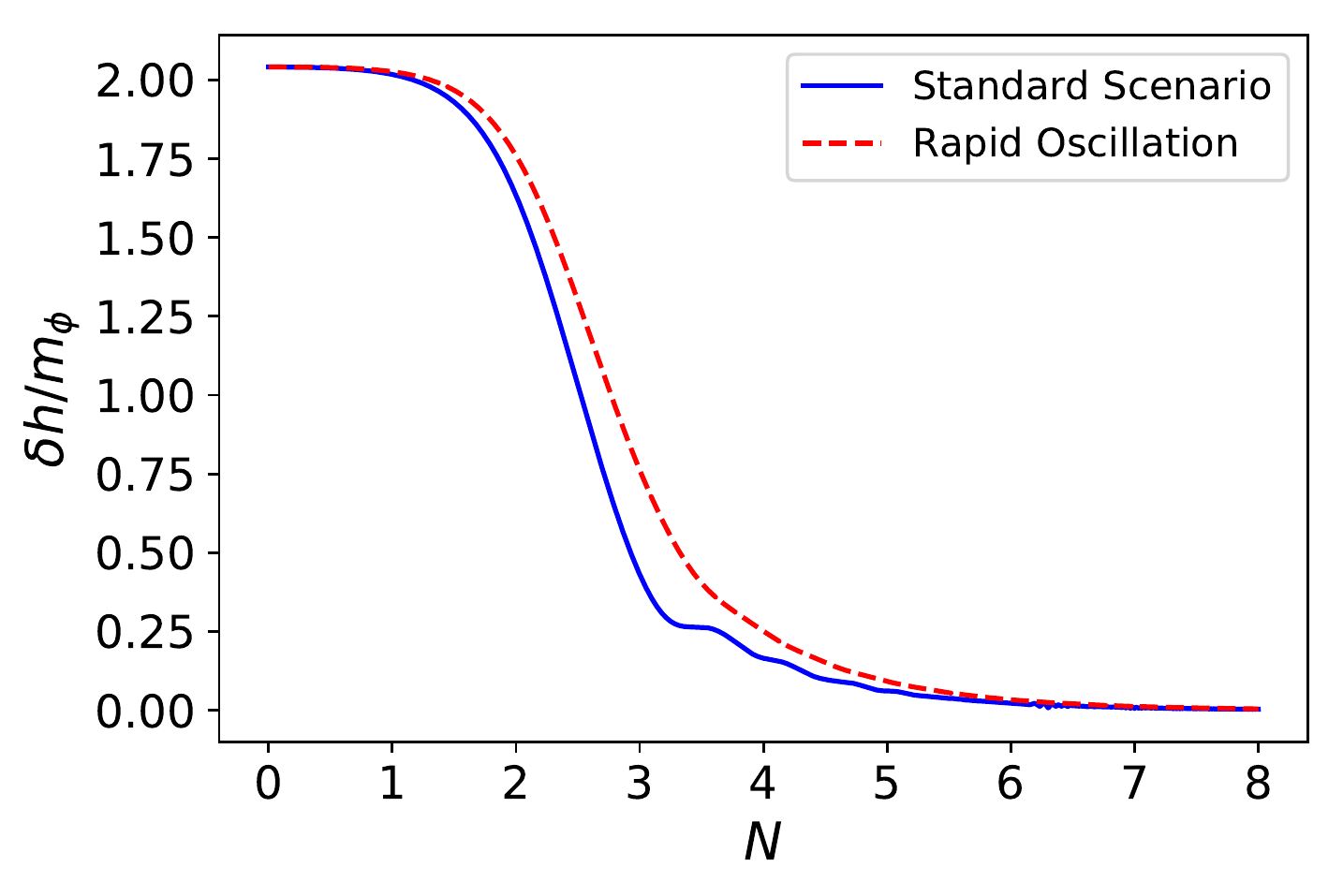} 
	\caption{For the case of perturbative inflaton decay, time-evolution of the Higgs PDF width $\delta h$ assuming rapid oscillations (red dashed line, $\THi \ll H^{-1}$) and our standard scenario assuming slow oscillations (blue solid line, $\THi \gg H^{-1}$). In the case of the former, the width corresponds to the standard deviation of a PDF formed by the effective Higgs values $h_{\rm eff}$ defined in Eq.~\eqref{eq:define_heff} and is given by $\delta h_{\rm rapid} = \sqrt{\langle h_{\rm eff}^2\rangle}$. In the case of the latter, the width is calculated using the PDF of Higgs VEVs $h$ presented in Fig.~\ref{fig:PDFhG}, as $\delta h_{\rm slow} = \sqrt{\langle h^2\rangle}$. We use $H_I=m_{\phi}$, $y = 1$, $\Gamma_0/m_{\phi} = 0.1$ and $\lambda_I=10^{-3}$.}
	\label{fig:deltah} 
\end{figure}

If, however, many Higgs oscillations occur during one Hubble timescale $\THi \ll H^{-1}$, the oscillations cannot be accurately resolved. We thus define an effective value of the Higgs field as 
\begin{eqnarray}
	\label{eq:define_heff}
	h^j_{\rm eff}(t) \equiv \rho_{h^j}^{1/4} = \left[\frac{1}{2}\left(\dot{h}^j\right)^2+V_H(h^j)\right]^{1/4}\;,
\end{eqnarray}
in a patch denoted by $j$, where $\rho_{h^j}$ is the energy density in the Higgs condensate at the particular patch. A comparison between the time-evolution of the Higgs PDF width $\delta h$ under the assumption of rapid oscillations ($\THi \ll H^{-1}$) and our standard scenario (where $\THi \gg H^{-1}$) is shown in Fig.~\ref{fig:deltah}. For rapid oscillations, the width corresponds to the standard deviation of a PDF of effective Higgs values $h_{\rm eff}$, as defined in Eq.~\eqref{eq:define_heff}, and is given by $\delta h_{\rm rapid} = \sqrt{\langle h_{\rm eff}^2\rangle}$. In our standard scenario, the width is calculated using the PDF of Higgs VEVs $h$ presented in Fig.~\ref{fig:PDFhG}, as $\delta h_{\rm slow} = \sqrt{\langle h^2\rangle}$. Since $\langle{h}\rangle = \langle{h_{\rm eff}}\rangle = 0$ for a symmetrical Higgs distribution, this figure also describes the evolution of the characteristic Higgs values $\tilde{h} = \langle{h}\rangle + \delta h_{\rm slow}$ and $\tilde{h}_{\rm eff} = \langle{h_{\rm eff}}\rangle + \delta h_{\rm rapid}$, which governs both $\bar{\Gamma}_{\phi}$ and $\delta_{\Gamma}$.  We can see that in the rapid oscillation case $\delta h_{\rm rapid}$ and hence $\tilde h_{\rm eff}$ decrease more slowly. As a result, Higgs blocking is lifted somewhat later under the assumption of rapid Higgs oscillations when compared to the scenario studied in Sec.~\ref{subsec:PbPHiggs} with $\THi \gg H^{-1}$.

\begin{figure*}
	\begin{center}
		\includegraphics[width=1.0\linewidth]{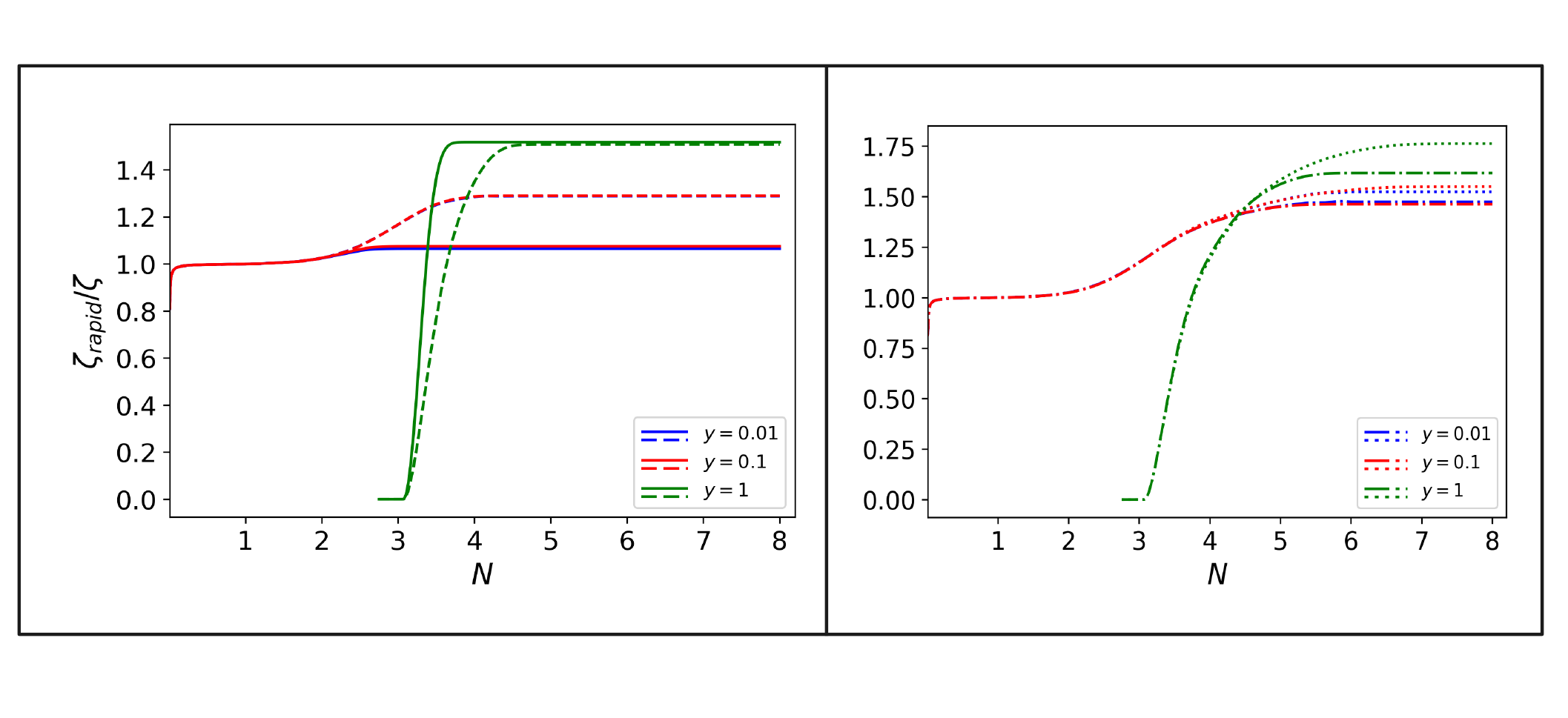}
		\caption{For the case of perturbative inflaton decay: The ratio of the Bardeen parameter $\zeta_{\text{rapid}}(N)$ calculated under the assumption of rapid Higgs oscillations ($\THi \ll H^{-1}$) to the Bardeen parameter $\zeta$ in the right panel of Fig.~\ref{fig1}, calculated assuming slow oscillations ($\THi \gg H^{-1}$). Results are shown as a function of the number of $e$-folds $N$ after inflation for three different values of the Yukawa coupling, $y=0.01$ (blue), $0.1$ (red), and $1$ (green). Different line styles correspond to different values of the unblocked decay rate $\Gamma_0 = 0.1\,m_{\phi}$ (solid, left panel), $\Gamma_0 = 0.01\,m_{\phi}$ (dashed, left panel), $\Gamma_0 = 10^{-3}\,m_{\phi}$ (dash-dot, right panel) and $\Gamma_0 = 10^{-4}\,m_{\phi}$ (dotted, right panel). We have taken $H_I = m_{\phi}$ and $\lambda_I = 10^{-3}$.   The ratio of Bardeen parameters is at most ${\cal O}(1)$, implying good agreement between the results of the two regimes. Thus, we can trust that the calculation of perturbations under the more realistic assumption of $\THi \simeq H^{-1}$ would yield similar results.}
		\label{fig4} 
	\end{center}
\end{figure*}

More typically, the Higgs oscillations would take place in the intermediate regime $\THi \simeq H^{-1}$, so that the actual perturbation values lie between the results derived in the two limiting scenarios mentioned above. In Fig.~\ref{fig4} we present a comparison between results derived in the two regimes for Yukawa couplings $y = 0.01$, $y = 0.1$ and $y=1$. Similarly to the backreaction comparison, we plot the ratio of Bardeen parameters calculated in the rapid Higgs oscillation scenario ($\THi \ll H^{-1}$) to those shown in Fig.~\ref{fig1} for the slowly-oscillating Higgs. For the case of $y=1$ (green lines), the initial value of the ratio begins at $0$ instead of $1$ due to the effect of Higgs blocking\footnote{The reason that such an effect is only visible for $y = 1$ is that this is the only case in which inflaton decay is actually blocked by the Higgs boson. In other words, this is the only case in which the argument of the Heaviside function in Eq.~\eqref{eq:decayrate} ever becomes negative.} which is effective for a longer time when assuming rapid Higgs oscillations. Also when assuming $\THi \ll H^{-1}$, sampling $h_{\rm eff}$ instead of $h$ causes Higgs blocking to be lifted slightly later, as explained above. As a result, $\zeta$ becomes non-zero, while $\zeta_{\rm rapid}$ is still blocked, thus leading to the ratio $\zeta_{\text{rapid}}/\zeta$ smoothly increasing from zero. 

We also observe in Fig.~\ref{fig4} that the importance of how we treat the Higgs oscillations is largely dependent on the decay rate of the inflaton. More specifically, smaller decay rates amplify the differences between the perturbations calculated in standard and rapid oscillation scenarios because reheating takes place later. As a result, for lower $\Gamma_0$ the Bardeen parameters increase the most between $N=2$ and $N=5$ $e$-folds, during which the widths $\delta h$ differ the most between the two cases (cf. Fig~\ref{fig:deltah}). If, on the other hand, the decay rate is larger then the Bardeen parameters increase most during the first $2$ $e$-folds, when $\delta h$ is the same in both cases. Therefore, the differences between the associated perturbations are smaller for larger $\Gamma_0$. 

More generally, we find that the Bardeen parameters calculated assuming rapid Higgs oscillations ($\THi \ll H^{-1}$) do not differ from the standard case ($\THi \gg H^{-1}$) by more than $ \mathcal{O}(1)$ factors; i.e. $\zeta_{\text{rapid}}/\zeta \sim 1$ in Fig.~\ref{fig4}. Since, as mentioned above, the Bardeen parameters calculated under the more realistic assumption of $\THi \simeq H^{-1}$ should lie in between those calculated in the two limiting cases, reasonably good agreement between both calculations allows us to confidently ascertain the corresponding constraints from temperature anisotropies in the CMB (see Sec.~\ref{sec:comparison}).

\begin{figure*}[t]
	\includegraphics[width=1\linewidth]{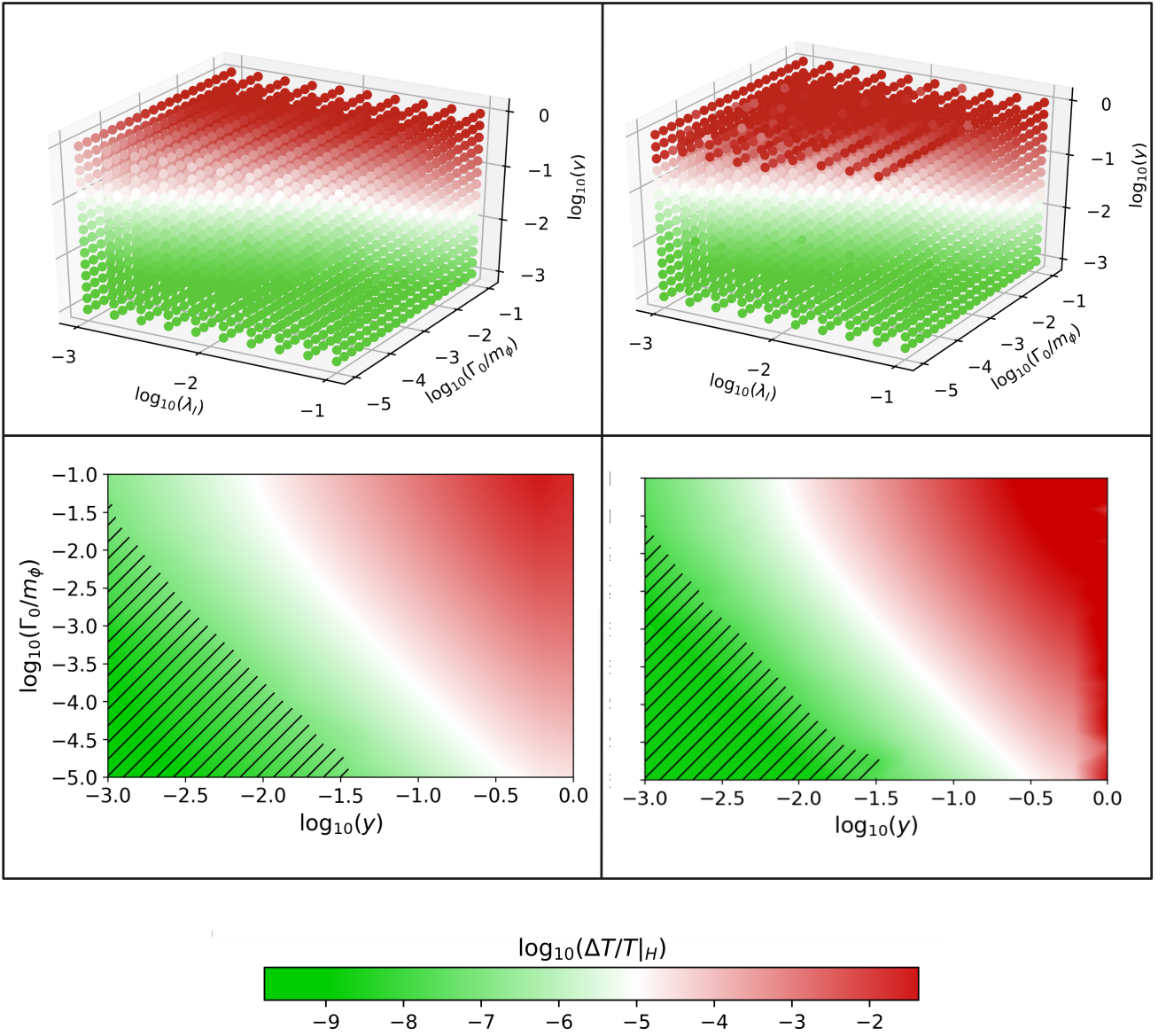} 
	\setlength{\abovecaptionskip}{-200pt}
	\caption{For the case of perturbative (space-dependent) inflaton decay to SM particles, constraints on parameters obtained by requiring that temperature fluctuations do not exceed CMB observations.  Here $\Gamma_0/m_\phi$ is the unblocked inflaton decay rate in units of inflaton mass; $y$ is the Yukawa coupling of the SM particles to the Higgs; and $\lambda_I$ is the Higgs self-coupling at the scale of inflation.  In the top panels of the figure, we depict the three-dimensional parameter space $\left(\lambda_I, y, \Gamma_0\right)$, and in the bottom panels we depict two-dimensional slices at the value $\lambda_I = 10^{-2}$, which is typical in the SM for inflation at high scales. The left (right) panels show results derived in the $\THi \gg H^{-1}$ ($\THi \ll H^{-1}$) regime of slow (rapid) Higgs oscillations. The red and green regions show Higgs-induced temperature inhomogeneities $\Delta T/T|_{\text{H}}$ which are larger and smaller than those observed in the CMB, respectively.  Hence the white and green regions are allowed as they satisfy $\Delta T/T|_{\text{H}} \leq 10 ^{-5}$.  The hatched region at $\Delta T/T|_{\text{H}} \leq 10 ^{-7}$ indicates the untested regime where the temperature fluctuations $\Delta T/T|_{\text{H}}$ are smaller than Planck's ${\cal O}(1)\%$ sensitivity (e.g. if we assume that the origin of the temperature anisotropies observed by Planck arise from the quantum fluctuations of the inflaton itself). For the intermediate regime of Higgs oscillations  with $\THi \simeq H^{-1}$, the amplitude of temperature fluctuations would lie somewhere in between the values shown for the slow- (left panels) and rapid-oscillation (right panels) approximations.}
	\label{fig5} 
\end{figure*}

\subsection{Comparison of Temperature Fluctuations from space-dependent Perturbative Reheating with CMB data}
\label{sec:comparison}

In this section we compare results of our calculations of the Bardeen parameter at the end of reheating with the value of the temperature anisotropy measured in the CMB, $\Delta T/T\big|_{\rm CMB} \approx 10^{-5}$. Since the predictions of the temperature fluctuations in our scenario depend on several parameters, we may use the CMB bounds to place constraints on the associated parameter space shown in Fig.~\ref{fig5}.  Specifically we examine the dependence of the temperature fluctuations on the Higgs self-coupling $\lambda_I$, the Yukawa coupling $y$ of the fermions to the Higgs, and the unblocked decay rate of the inflaton $\Gamma_0$. In the top panels of the figure, we depict the three-dimensional parameter space $\left(\lambda_I, y, \Gamma_0\right)$, and in the bottom panels we depict two-dimensional slices at the value $\lambda_I = 10^{-2}$ which is typical for the SM in high scale inflation. We explore a wide range of different values for the unblocked decay rate $10^{-5}\le \Gamma_0/m_\phi \le 10^{-1}$ and the Yukawa coupling $10^{-3}\le y \le 1$.  

In all diagrams, red regions correspond to parameter combinations that lead to Higgs-modulated fluctuations in excess of what is observed in the CMB ($\Delta T/T|_{\text{H}} \gtrsim \Delta T/T\big|_{\rm CMB}$); and green regions correspond to Higgs-induced fluctuations that are below the CMB values ($\Delta T/T|_{\text{H}} \lesssim \Delta T/T\big|_{\rm CMB}$). The white region indicates Higgs-induced fluctuations of the same size as those of CMB observations $\Delta T/T\big|_{\text{H}} \approx 10^{-5}$. Hence both white and green regions are in principle allowed by the CMB, while the red region is certainly excluded.

We assume CMB observations are not sensitive to Higgs-induced perturbations with $\Delta T/T|_{\text{H}} \lesssim 0.01\,\Delta T/T\big|_{\rm CMB} \sim 10^{-7}$. The reason is that such small temperature anisotropies are below the $\mathcal{O}(1\%)$ sensitivity of Planck. Then the observed anisotropies of the CMB $\Delta T/T\big|_{\rm CMB} \approx 10^{-5}$ must be produced via some other mechanism, e.g. the standard density fluctuations arising from quantum fluctuations of the inflaton field. The regime for the Higgs-induced fluctuations $\Delta T/T|_{\text{H}} \lesssim 10^{-7}$ is indicated by the hatched region of Fig.~\ref{fig5}. 

The left (right) panels of Fig.~\ref{fig5} are for the cases of slow (rapid) Higgs oscillations. Sampling the effective Higgs values of Eq.~\eqref{eq:define_heff} (rapid oscillations, right panels of Fig.~\ref{fig5}) leads to slightly larger overall perturbations than the case of slow oscillations. However, one can see by comparing the left and right panels that the observed differences are very small and much below an order of magnitude. As explained in Sec.~\ref{sec:Assuming time-averaged Higgs oscillations}, the true period of the Higgs oscillations could in fact be in the intermediate regime between the two limiting cases we investigate, with $\THi \simeq H^{-1}$. In the intermediate case, the  amplitude of temperature fluctuations would lie in between those calculated in the slow- and rapid-oscillation approximations, which are in reasonably close agreement.

Let us now summarize the basic parameter dependencies of Fig.~\ref{fig5}. The Higgs quartic self-coupling $\lambda_I$ at the end of inflation only appears in the three-dimensional plots in the top panels of Fig.~\ref{fig5}. It is evident that the size of temperature fluctuations decreases as $\lambda_I$ increases from $10^{-3}$ to $10^{-1}$. This effect can be explained by the width of the Higgs PDF at the end of inflation, given by Eq.~\eqref{eqII8}, decreasing for larger $\lambda_I$. In other words, Higgs PDFs with a larger value of $\lambda_I$ are peaked around smaller values of the Higgs field $h$ and give smaller probabilities for larger values of $h$ to exist in a given Hubble patch. Since the Higgs modulation that causes the temperature fluctuations is more pronounced for larger values of the Higgs field $h$, a larger $\lambda_I$ leads to smaller overall $\Delta T/T|_{\rm H}$. 

The unblocked decay rate $\Gamma_0$ of the inflaton to fermions and of the Yukawa coupling $y$ of the same fermions to the Higgs appear in both the three- and the two-dimensional plots of Fig.~\ref{fig5}. We can see larger overall temperature fluctuations arise for larger $y$ and for larger $\Gamma_0$. Larger $y$ leads to larger fermion masses and, as a result, to more significant Higgs modulation of the decay rate. Since Higgs modulation is the seed of the temperature fluctuations we examine, larger $y$ will also lead to larger overall temperature fluctuations. The effects of increasing the decay rate are more subtle, but can be understood from Fig.~\ref{fig1}. The inflaton decays at earlier times in each Hubble patch for larger $\Gamma_0$. Values of the Higgs VEV in each patch, evolving according to Eq.~\ref{eq:dhdN}, are larger at earlier times. Thus, larger $\Gamma_0$ leads to larger fermion masses when the inflaton decays, also causing larger overall temperature fluctuations.

Here we note several caveats in our calculations due to some of the other approximations we have made. First, when deriving the perturbations we only treat the largest observable scales in the CMB. This simplification is inevitable since our method uses the one-point PDF of the Higgs field, which lacks any information regarding scale-dependence. Therefore, our results are consistent with the assumption of a scale-invariant power-spectrum of Higgs fluctuations and remain a very good approximation (up to factors of $\sim \mathcal{O}(1)$) for mild scale dependencies. 

In a related approximation mentioned in previous sections, our calculations assume a purely de-Sitter spacetime when the largest observable scales exit the horizon during inflation and that the PDF of the superhorizon modes of the spectator Higgs field has reached its equilibrium. Without this assumption we would not be able to sample the equilibrium distribution function of Eq.~\eqref{eqII8} for the initial condition of the Higgs in each Hubble patch. We leave a more detailed treatment of the Higgs fluctuations for future work, noting in particular that calculations valid at smaller angular scales could allow for constraints in the (unhatched) green region of Fig.~\ref{fig5}. 

In order to more clearly demonstrate the parameter dependence of the temperature fluctuations $\Delta T/T|_{\text{H}}$ produced by Higgs modulated reheating, we estimate a fitting function in terms of the Yukawa coupling $y$ and the unblocked decay rate $\Gamma_0$ based on parameter space data shown in the bottom left panel of Fig.~\ref{fig5}, 
\begin{equation}
\label{eq:fittedfunc}
\begin{split}
	\log_{10}\left[\frac{\Delta T}{T}\bigg|_{\text{H}}(y, \Gamma_0)\right] &= 1.41\log_{10}(y)-0.14\left[\log_{10}(y)\right]^2\\
	&-0.07\left[\log_{10}(\Gamma_0/m_{\phi})\right]^2 - 0.8\\
	\\
	&-0.05\log_{10}(y)\log_{10}(\Gamma_0/m_{\phi})\\
	\\
	&+0.37\log_{10}(\Gamma_0/m_{\phi})\;.
\end{split}
\end{equation}

Using this fitting function, we can constrain parameter values\footnote{The values mentioned are subjected to the error of our fit. For more accuracy see Figs.~\ref{fig1} and \ref{fig5}.} for each SM particle separately based on its Yukawa coupling, provided that the inflaton decay channel we are examining is the dominant one during reheating. As a reminder, the inflaton can always decay into massless photons without the Higgs influencing the size of the produced temperature fluctuations. Furthermore, while there is a running of the fermionic Yukawa couplings between the Electroweak and the inflation scale, we do not expect them to be significantly modified and, thus, we choose to use their standard EW values.

We compare the Higgs-induced temperature fluctuations (calculated from Eq.~\eqref{eq:fittedfunc}) produced for different Yukawa couplings with the CMB value of $\Delta T/T|_{\rm CMB}\approx 10^{-5}$ and place upper bounds on the quantity $\Gamma_0 / m_\phi$ where $\Gamma_0$ is the unblocked decay rate. For the cases with Yukawa coupling $y<1$, we previously showed in FSSV that Higgs blocking is minimal and we can approximate the time of reheating by $\Gamma_0\sim H$. The reheat temperature $T_{\text{reh}}$ is then given directly in terms of $\Gamma_0$ as 
\begin{equation}
\label{eq:Treh}
T_{\text{reh}} = \left(\frac{5}{4\pi ^3 g_*}\right)^{1/4}\sqrt{\Gamma_0\, m_{\text{Pl}}}\approx 0.14\left(\frac{100}{g_*}\right)^{1/4}\sqrt{\Gamma_0\, m_{\text{Pl}}}\, 
\end{equation}
where $g_*\approx 106.75$ is the number of relativistic degrees of freedom of the radiation bath. The same relation applies even for cases with $y = 1$ where Higgs blocking is present, provided that the unblocked decay rate $\Gamma_0$ is sufficiently low. This condition ensures that blocking will be lifted before $\Gamma_0\sim H$. When this condition is not fulfilled, Eq.~\eqref{eq:Treh} gives a reheat temperature larger than its actual value since it does not take the delays due to Higgs blocking into consideration. In Sec.~\ref{sec:Simplest_perturbative_reheating_case} we explained how Fig.~\ref{fig1} shows that for $y=1$ and $\Gamma_0/m_{\phi}\lesssim 10^{-2}$ Higgs blocking is lifted before $\Gamma_0\sim H$. In the following we will see that our upper bound for $\Gamma_0$ in the case of the top-quark ($y_t = 1$) is $\Gamma_0/m_{\phi}|^{\text{max}}_{\text{top}} \ll 10^{-2}$ and, thus, Eq.~\eqref{eq:Treh} still applies.\footnote{Sec.~\ref{sec:Simplest_perturbative_reheating_case} uses $\lambda_I = 10^{-3}$, whereas here we use $\lambda_I = 10^{-2}$. However, for larger values of the self-coupling $\lambda_I>10^{-3}$ blocking is lifted even sooner and, thus, our statement remains accurate. Additionally we have made sure that blocking is lifted before $\Gamma_0\sim H$ for $\Gamma_0 = \Gamma_0/m_{\phi}|^{\text{max}}_{\text{top}}$ numerically.} Hence, by setting upper bounds on the decay rate $\Gamma_0$ we can immediately constrain the reheat temperature for inflaton decays to all SM fermions. 

We assume the maximum allowed value of the unblocked decay rate to be $\Gamma_0 \lesssim 10^{-1}\times m_{\phi}$ in order to ensure that the decay of the inflaton does not arise from a strongly coupled theory. A corresponding lower bound does not exist, provided that reheating is complete by the time of Electroweak Symmetry breaking (EWSB) or, at the latest, Big Bang Nucleosynthesis (BBN), whose energy scale is much lower than the inflationary one \cite{Hasegawa:2019jsa}. According to Eq.~\eqref{eq:Treh}, the temperature of EWSB $T_{\text{EWSB}}\approx 160$ GeV corresponds to $\Gamma_0|_{\text{EWSB}} \approx 10^{-24}\times m_{\phi}$, which is much below any of the limits we are setting. Stronger lower bounds on $T_{\rm reh}$ would apply if we wanted (vanilla) thermal leptogenesis to provide the origin of today's matter/antimatter asymmetry; then the bounds would be $T_{\rm reh} > 10^{9-11}$ GeV depending on the model (for example, see Ref.~\cite{Giudice:2003jh}).

Let us consider the potentially most constrained case, i.e., where the inflaton primarily decays to the top quark, the most massive of the fermions with the largest SM Yukawa coupling $y_t\approx 1$. We find that the upper bound on the decay rate is $\Gamma_0/m_{\phi}|_{\text{top}} < 3\times 10^{-6}$ corresponding to a bound on the reheat temperature 
\begin{equation}
	T_{\text{reh}}|_{\text{top}} < 2.4\times 10^{12} {\rm \, GeV} \,\, \biggl({m_\phi \over 10^{13} {\rm \, GeV}} \biggr)^{1/2}  .  
\end{equation}
This bound may be in conflict with models of thermal leptogenesis,  depending on the mass of the inflaton and the lightest right-handed neutrino, but will never be in conflict with EWSB or BBN.  

For the electron and the muon, with $y_e \approx 3\times 10^{-6}$ and $y_{\mu} \approx 6\times 10^{-4}$ respectively, any value of the unblocked decay rate yields temperature fluctuations which are unconstrained by CMB observations and lie within the hatched region of Fig.~\ref{fig5} (bottom panels). For inflaton decays to the the tau lepton, with $y_{\tau} \approx 10^{-2}$, we find the constraint $\Gamma_0/m_{\phi}|\tau < 0.04$ corresponding to 
\begin{equation}
T_{\text{reh}}|_{\tau} < 2.8\times 10^{14} {\rm \, GeV} \biggl({m_\phi \over 10^{13} {\rm \,GeV}} \biggr)^{1/2}. 
\end{equation}

\subsection{Lowering the scale of inflation}
\label{sec:lowerscale}

\begin{table*}[tb]
	\begin{tabular}{|c||c|c|c|c|c||}
		\cline{2-6}
		\multicolumn{1}{l|}{}                 & High-Scale Inflation          & \multicolumn{4}{c|}{Lowering the Scale of Inflation}                                                                                      \\ \hline\hline
		\multicolumn{1}{|c||}{Inflation Scale} & $H_I =m_{\phi}$               & $H_I =10^{-1}\, m_{\phi}$ & $H_I =10^{-2}\, m_{\phi}$ & $H_I =10^{-3}\, m_{\phi}$ & $H_I =10^{-4}\, m_{\phi}$ \\ \hline
		\multicolumn{1}{|c||}{Decay Rate}      & $\Gamma_0/m_{\phi} = 10^{-1}$ & $\Gamma_0/m_{\phi} = 10^{-2}$ & $\Gamma_0/m_{\phi} = 10^{-3}$ & $\Gamma_0/m_{\phi} = 10^{-4}$ & $\Gamma_0/m_{\phi} = 10^{-5}$ \\ \hline
		\multicolumn{1}{|c||}{Yukawa Coupling} & $y = 10^{-2}$          & $y =10^{-1}$          & $y = 1$          & $y =10$                        & $y = 100$                      \\ \hline
	\end{tabular}
	\caption{For the case of perturbative inflaton decay, values of the Yukawa couplings $y$ above which parameter space is excluded for certain choices of unblocked decay rate $\Gamma_0/m_{\phi}$ and inflation scale $H_I/m_{\phi}$. When considering a lowered inflationary scale, for simplicity we keep the value of the  inflaton mass $m_{\phi}$ equal to that in the case of high-scale inflation. We choose $\lambda_I = 10^{-2}$.}
	\label{tab1}
\end{table*}

In the previous sections we have explored the temperature fluctuations arising from Higgs modulation/blocking of reheating after high-scale inflation, by setting the Hubble parameter at the end of the inflationary epoch equal to the mass of the inflaton $H_I = m_{\phi}$. However, it is also useful to look into cases where the Hubble parameter $H_I$ takes much smaller values and, thus, investigate what our perturbations would look like if inflation were to occur at a lower scale.

Upon examination of the equations which govern the production of the gravitational and Bardeen parameters, as well as numerically solving the same system of equations for various different combinations of the input parameters, we see that inflation at a lower scale can produce the same perturbations as those of high-scale inflation. Re-scaling of our results for the amplitude of temperature fluctuations is possible provided that the free parameters are chosen such that 
\begin{eqnarray}
	\left[\Gamma_0\, H_I \, y^2 \right]_{\text{ls}} = \left(\frac{m_{\phi}|_{\text{ls}}}{m_{\phi}|_{\text{hs}}}\right)^2\left[\Gamma_0\, H_I \, y^2 \right]_{\text{hs}}\;,
	\label{eqIII5}
\end{eqnarray}
where the subscripts ``\textrm{hs}'' and ``\textrm{ls}'' refer to high and lower-scale inflation respectively. Another necessary condition for the applicability of our results to inflation at lower scales is that the unblocked decay rate $\Gamma_0$ should be smaller than the value of the Hubble parameter $H_I$ by the same amount as for inflation at high scales,
\begin{equation}
\left[\frac{\Gamma_0}{H_I}\right]_{\text{ls}} = \left[\frac{\Gamma_0}{H_I}\right]_{\text{hs}}\;.
\label{eqIII6}
\end{equation}
Inserting Eq.~\eqref{eqIII6} into Eq.~\eqref{eqIII5} gives
\begin{equation}
\left[H_I \, y\right]_{\text{ls}} = \frac{m_{\phi}|_{\text{ls}}}{m_{\phi}|_{\text{hs}}}\left[H_I \, y\right]_{\text{hs}}\;,
\label{eqIII7}
\end{equation}
and, therefore, Eqs.~\eqref{eqIII6} and \eqref{eqIII7} are the two independent conditions for high-scale and lower-scale inflation to result in the same perturbations.

The re-scaling of the results of our paper as a function of inflation scale can be understood by examining the role of these two conditions. Going back to Eqs.~\eqref{eq:Boltzmann_mN}-\eqref{eq:Boltzmann_rN} for the background evolution, we see that the factor setting different Hubble patches apart is
\begin{equation}
 {\bar{\Gamma}_{\phi} \over H_I } = \left(\frac{\Gamma_0}{H_I}\right)\left(1-\frac{2y^2 h^2 }{m_{\phi}^2}\right)^{3/2}\Theta\left(m_{\phi}^2 - 2y^2 h^2\right)\;.
\label{eqIII8}
\end{equation}
We notice that the two important quantities $\Gamma_0/H_I$ and $y\, h/m_{\phi}\propto y\, H_I/m_{\phi}$ are the same regardless of the inflationary scale if the parameters are chosen such that they match the conditions in Eqs.~\eqref{eqIII6}-\eqref{eqIII7}. In Eqs.~\eqref{eq:delta_Hubble_N}-\eqref{eq:delta_Boltzmann_m_N}, which govern the growth of perturbations in our model, the term of importance which essentially \textit{fuels} the potential production, is 
\begin{equation}
	\label{eqIII9}
	 \frac{\bar{\Gamma}_{\phi}}{H_I} \, \delta_{\Gamma_{\phi}}\propto \frac{\Gamma_0 H_I}{m_{\phi}^2} y^2 \left(1-\frac{2y^2 H_I^2 }{m_{\phi}^2}\right)^{1/2}\Theta\left(m_{\phi}^2 - 2y^2 H_I^2\right)\;,
\end{equation}
for $h\propto H_I$. Again, we notice that the factor $\Gamma_0 H_I y^2/m_{\phi}^2$ is set to be the same regardless of the inflationary scale if we choose parameters satisfying Eq.~\eqref{eqIII5}, while the term $y\, H_I/m_{\phi}$ is determined by Eq.~\eqref{eqIII7} for appropriately chosen parameters. 

We, therefore, conclude that a lower inflation scale $H_I$ will lead to a re-scaling of the parameter space introduced in Fig.~\ref{fig5} and allow larger values of the Yukawa couplings $y$. In other words, the green region of the same Figure will move further up to larger values of $\log(y)$ and our constraints will be relaxed.

Table~\ref{tab1} presents the largest Yukawa couplings allowed by observations for certain choices of the unblocked decay rate $\Gamma_0$ and of the inflation scale $H_I$. We can see that as the scale of inflation goes down the allowed values of $y$ increase, provided that $\Gamma_0$ is chosen such that Eq.~\eqref{eqIII6} is satisfied. For example, for inflation scale $H_I = 10^{-4} m_\phi$ (four orders of magnitude lower than our canonical case), and taking decay rate $\Gamma_0/m_\phi = 10^{-5}$ to match the condition in Eq.~\eqref{eqIII6}, the Yukawa coupling $y$ can be as large as 100 without violating CMB constraints.\footnote{While values of the Yukawa coupling $y \gtrsim 4 \pi$ are larger than what is typically permitted by perturbative unitarity, as discussed in FSSV the relevant parameter is actually the mass of the fermion given by Eq.~\eqref{eq:fermionmass} and, thus, such large Yukawa couplings can be considered representative of cases with $y \lesssim 4 \pi$ and proportionally larger Higgs field values (cf. discussion at the end of Sec.~\ref{sec:Simplest_perturbative_reheating_case}).} In order to derive these results we have taken $m_{\phi}$ to be the same both in high-scale and in the case of a lower inflationary scale, even though it could in principle vary between the two. 

A commonly used example in the literature for constructing models with a low inflationary scale consistent with the CMB are $\alpha$-attractor models, like the T-model~\cite{Kallosh:2013yoa}, whose single-field potential is
\begin{equation}
	\label{eq:alphaattractor}
	V(\phi) = \mu^2 \alpha \tanh^2 \left(\phi \over \sqrt{6 \alpha}\right)
\end{equation}
The mass scale $\mu = {\cal O}\left(10^{-6}\right)\,M_{\rm Pl}$ is chosen to set the amplitude of curvature perturbations $\zeta \sim 10^{-5}$, in line with the results of the \textit{Planck} collaboration~\cite{Akrami:2018odb}. The parameter $\alpha$ controls the tensor-to-scalar ratio and the scale of inflation, which scales as $H\propto \sqrt{\alpha}$. This potential  shows a simple realization of a model where lowering the inflationary scale does not affect the inflaton mass close to the minimum of the potential, $m_\phi^2 = \mu^2/3$.  It has nevertheless been shown that the T-model at low $\alpha$ preheats efficiently both through self-resonance and through parametric resonance of a companion spectator field~\cite{Lozanov:2017hjm, Iarygina:2018kee, Krajewski:2018moi}. 

In general, keeping a large inflaton mass while lowering the energy scale of inflation will likely introduce significant non-linear terms to the potential that will lead to significant self-resonance. This is required, since low-scale single-field slow-roll inflation requires a flat ``plateau'' in the potential and the transition from a large inflaton mass at low field values to a flat plateau at large field values requires non-linear terms in the potential. Keeping this in mind, a more thorough investigation of Higgs blocking effects in low-scale inflationary models must take self-resonance into account. Assuming that self-resonance occurs efficiently, the Universe at the end of low-scale inflation will be populated by scalar (inflaton) particles whose mass will be much larger than the Hubble scale, at least for the $\alpha$-attractor potential of Eq.~\eqref{eq:alphaattractor}. The momentum of these particles will be of the order of their mass, thus they will be either non-relativistic or slightly relativistic. We leave a detailed (and somewhat model-dependent) analysis of Higgs-modulated reheating effects in low-scale inflation for future work.

\section{Results for the Case of Resonant Inflaton Decay}
\label{sec:gauge}

\begin{figure}[h!]
	\includegraphics[width=1\linewidth]{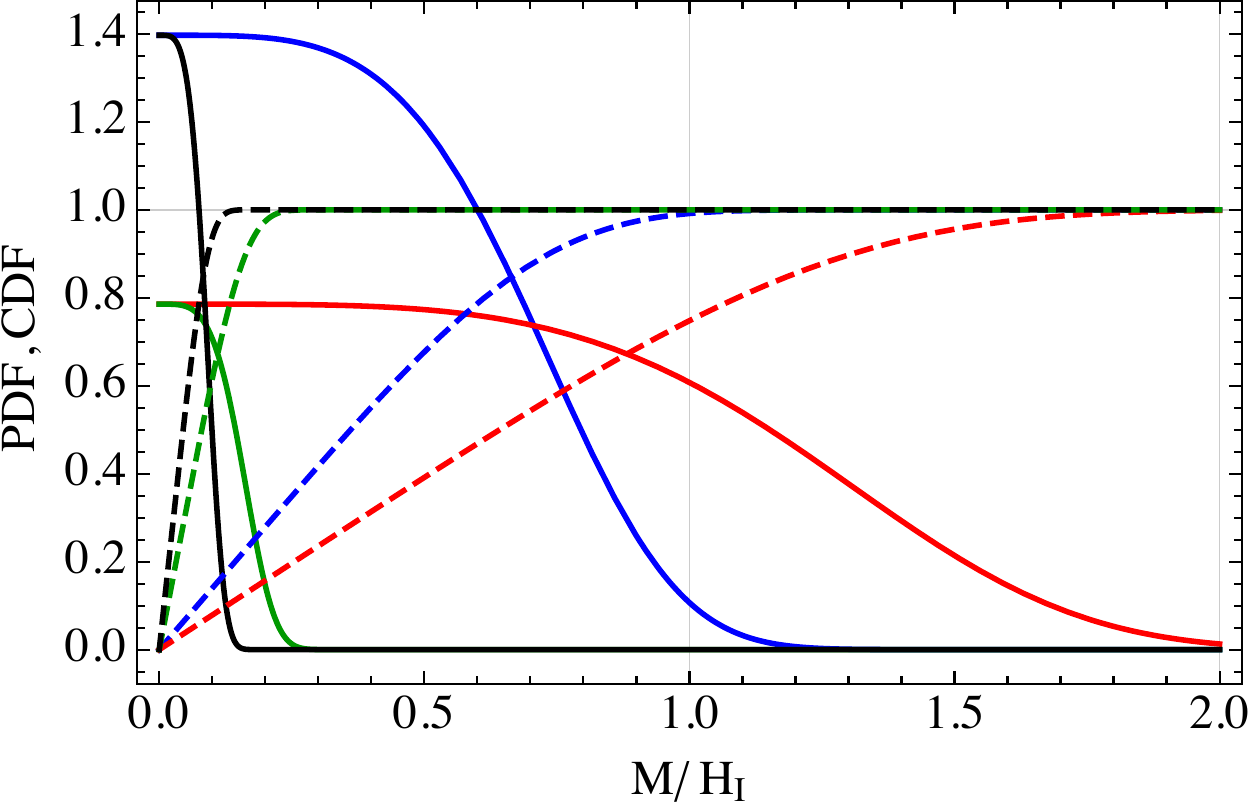} 
	\includegraphics[width=1\linewidth]{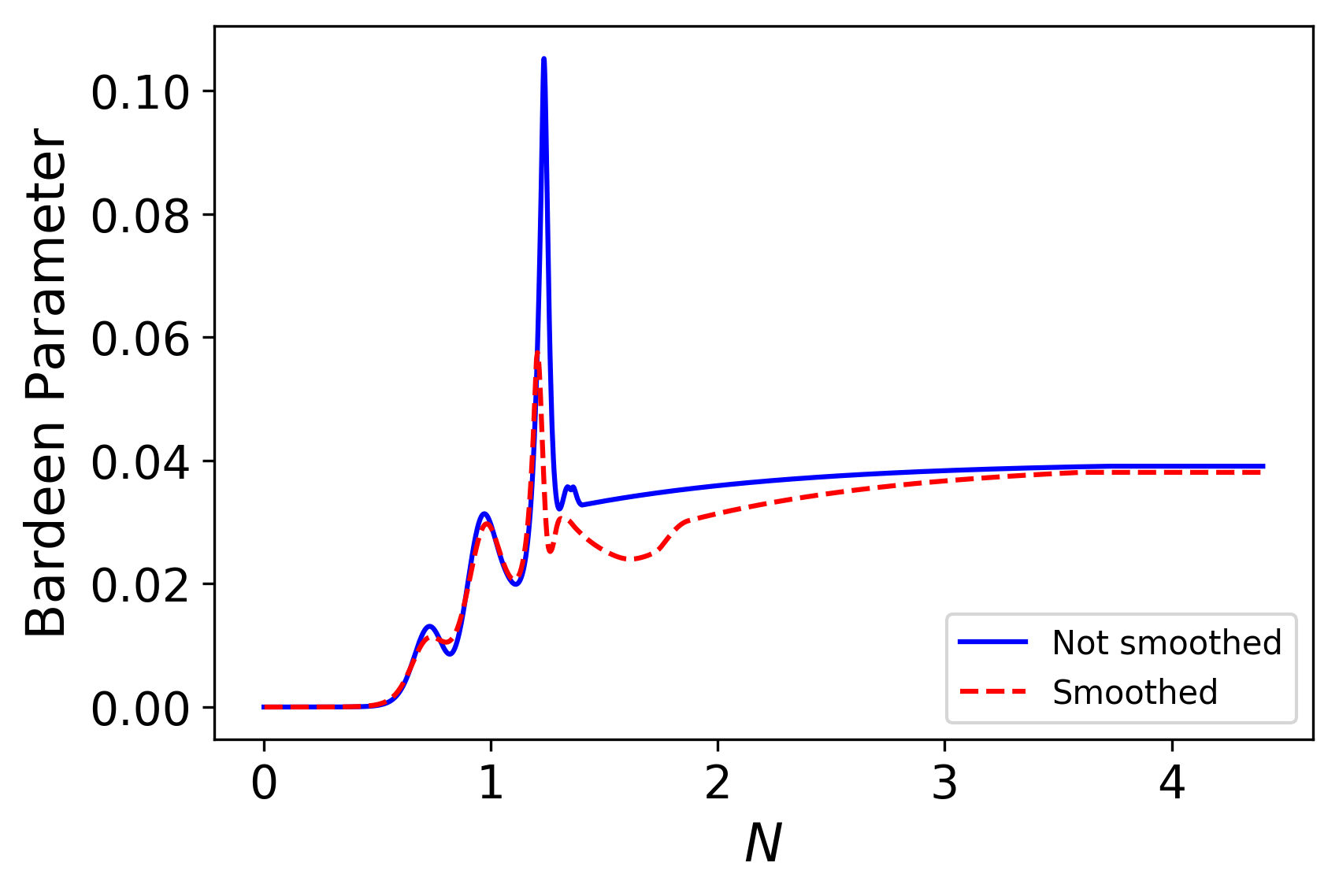} 
	\includegraphics[width=1\linewidth]{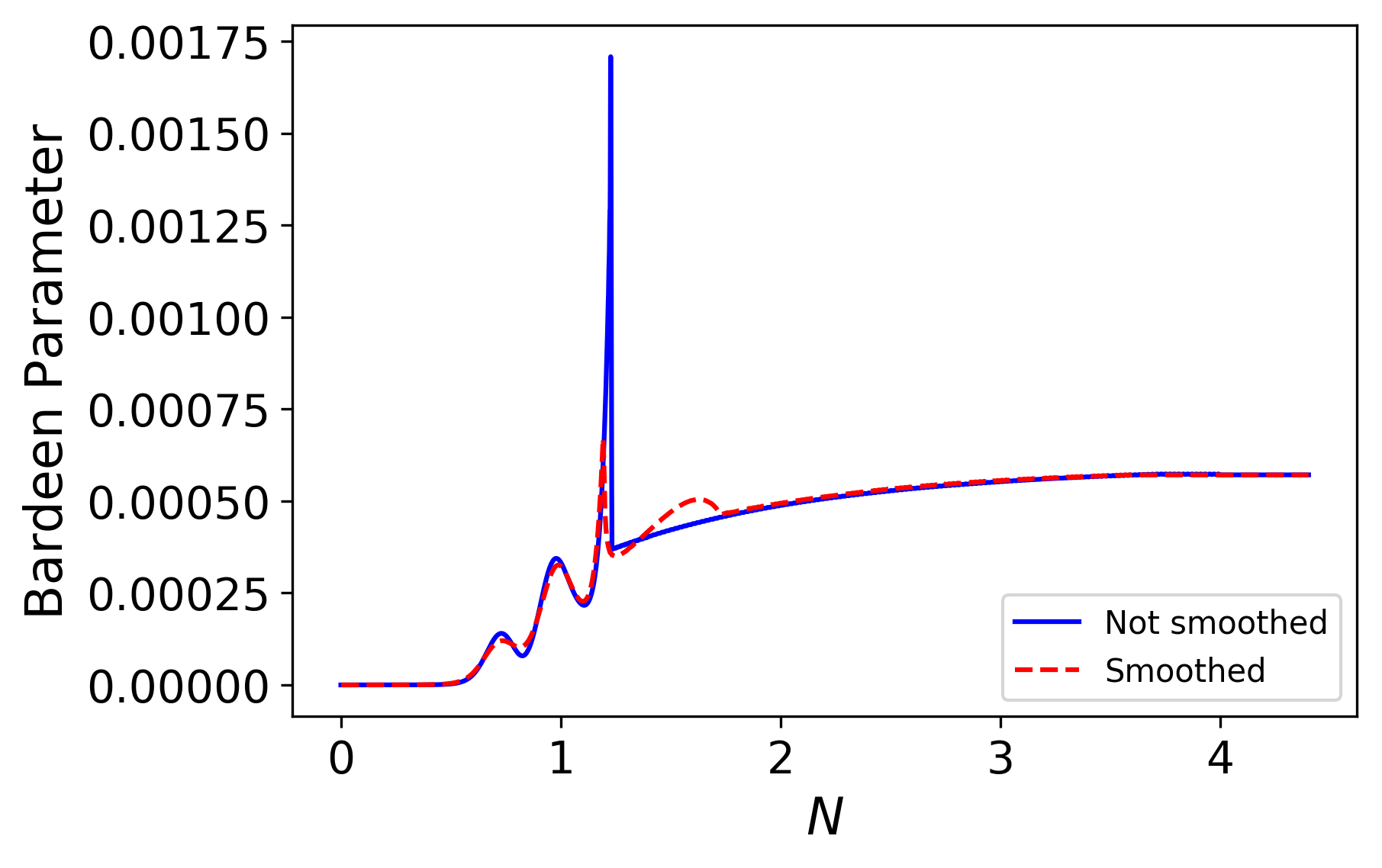} 
	\caption{Density peturbations generated by Higgs-modulated reheating in the case of resonant inflaton decay. 
		{\it Upper panel:} 
		The distribution of gauge field masses for $\{ g,\lambda\}=\{0.8, 10^{-3} \},\{0.8, 10^{-2} \},\{0.1, 10^{-3} \},\{0.1, 10^{-2} \} $ (red, blue, green and black respectively). The solid curves correspond to the PDF and the dashed curves correspond to the associated CDF. 
		The PDF for the mass is one-sided, since $m\propto |h|$, and thus the overall normalization differs from the expression which would be derived from Eq.~\eqref{eqII8} by a factor of $2$. 
		{\it Middle panel:} The Bardeen parameter $\zeta$ for perturbations generated by Higgs-modulated gauge preheating with $f=0.1\,m_{\rm Pl}$, $\lambda=10^{-2}$ and $g=0.8$.
		{\it Lower panel:} The corresponding Bardeen parameter $\zeta$ for $f=0.1\,m_{\rm Pl}$, $\lambda=10^{-2}$ and $g=0.1$.}
	\label{fig:gaugeresults} 
\end{figure}

We now present the effects of Higgs blocking and modulation on the generation of perturbations during gauge preheating, as a representative case of resonant decay of the inflaton. In FSSV we explored the effects of Higgs blocking on gauge preheating and found that complete preheating is only possible when the mass of the gauge bosons is $M\lesssim H_I$, with the exact threshold depending on the Chern-Simons coupling strength. The upper panel of Fig.~\ref{fig:gaugeresults} shows the probability distribution of gauge field masses $M=g |h| /2$ for two values of the gauge coupling $g=0.8$, $g=0.1$ and two values of the Higgs self-coupling $\lambda = 10^{-2}, 10^{-3}$, based on the probability distribution of Higgs values, given in Eq.~\eqref{eqII8}. 

For the case of $\lambda=10^{-3}$, there is a wider range of gauge field masses in different patches, reaching up to $M\simeq 2 H_I$ for $g=0.8$. Those Hubble patches in which the gauge field mass satisfies $M\lesssim H_I$ will successfully preheat resonantly; on the other hand those Hubble patches in which the gauge field masses are larger ($M>H_I)$ will not have successful resonant preheating and must reheat later through e.g. perturbative decay of the inflaton through the same coupling to gauge bosons. In the case of efficient preheating, the reheat temperature will be $T_{\rm reh}\sim \sqrt{m_{\rm Pl} H_I}$, while for perturbative decays the reheat temperature can be lowered by orders of magnitude~\cite{Freese:2017ace}. This range of behaviors from one Hubble patch to another will lead to large temperature fluctuations $\Delta T / T \sim 1$ for the case of $\lambda = 10^{-3}$. 

We thus restrict our analysis to $\lambda=10^{-2}$, where the gauge field masses predominantly satisfy $M\lesssim H_I$ (as seen in the upper panel of Fig.~\ref{fig:gaugeresults}), and assume the Chern-Simons coupling is large enough such that the entirety of the observable Universe can completely preheat through parametric resonance of gauge bosons. Although these assumptions prevent the generation of manifestly large temperature fluctuations associated with incomplete preheating in a significant number of Hubble patches, each patch will nevertheless preheat at a slightly different time. We thus explore how the distribution of gauge boson masses leads to inhomogeneous  preheating and determine if the amplitudes of the associated perturbations are in agreement with CMB observations. 

Following the discussion of Sec.~\ref{subsec:PbPendens} we solve Eq.~\eqref{eq:Ak}, the linearized evolution equation of the gauge field modes, for a grid of comoving wave numbers $k$, starting enough $e$-folds before the end of inflation, so that we can reliably initialize each mode in its Bunch-Davies vacuum state. The middle and lower panels of Fig.~\ref{fig:gaugeresults} show the results of using the linearized analysis of fluctuations until the point when the radiation energy density equals the inflaton background energy density in each patch. In order to extend the energy density evolution beyond the point of complete preheating, we assume that the entire energy density of each patch subsequently red-shifts like radiation, $\rho\propto a^{-4}$. This introduces a sharp ``knee'' around $N\simeq 1.2$ $e$-folds, which is evident in the evolution of the Bardeen parameter $\zeta$. 

In order to validate the approximation which assumes an instantaneous transfer of energy into radiation, we repeat the calculation of the Bardeen parameter but numerically smooth the transition from the inflaton dominated epoch to radiation domination. The evolution of the Bardeen parameters are shown using both approximations in the middle and lower panels of Fig.~\ref{fig:gaugeresults}. It is clear that the final amplitudes of the perturbations are largely independent of how we treat the transition between matter and radiation domination. The bulk of the evolution of the perturbations occurs within the first $e$-fold after inflation, when the gauge fields are amplified enough to account for a non-negligible part of the energy density of the Universe, at the percent level or above. Smoothing the transition between epochs does indeed make the evolution of $\zeta$ smoother, but altering the precise details of the energy densities will not significantly change the amplitudes of the perturbations. 

Furthermore, the middle and lower panels of Fig.~\ref{fig:gaugeresults} show that, in the case of parametric resonance, the effects of Higgs modulation lead to perturbations that are significantly larger than those observed in the CMB for both $g=0.8$ and $g=0.1$. Despite the uncertainty in the exact running of the SM couplings to the inflation scale, gauge couplings within the range $0.1\le g\le 0.8$ can represent a wide variety of different models. As a result, preheating into Higgsed gauge bosons cannot be the main source of reheating the Universe, at least given the fairly generic assumptions described in Sec.~\ref{sec:reheatProcesses}. If the inflaton couples to both a massless gauge boson (photon) and to massive ones ($W^\pm$ and $Z$), the problem becomes highly parameter dependent, as it depends crucially on the strength of the Chern-Simons coupling to each gauge boson.
We expect the effect to still be present in the full electroweak sector, but suppressed compared to our current analysis. A key factor in this process will be the ratio of the energy density of the inflaton that ends up in massive and massless gauge bosons. We leave a detailed computation of such a scenario for
 future work.

\section{Discussion and Conclusions} \label{sec:discussion}

If the Higgs field effective VEV has large non-zero fluctuations during inflation, it could imprint considerable effects on the subsequent stages of reheating, namely resonant particle production (preheating) and perturbative decays from coherent oscillations of the inflaton field. The quantum oscillations in the Higgs field give it a location-dependent effective VEV, imparting mass to any Standard Model (SM) particles to which it couples. If the particle mass exceeds the inflaton mass in some Hubble patch, then reheating there may be delayed, a phenomenon known as Higgs Blocking~\cite{Freese:2017ace}.

Adiabatic fluctuations arise because the Universe exhibits a space-dependent reheat temperature, 
due to the correspondingly space-dependent Higgs-induced particle masses. Consequently, density perturbations are created that later source CMB fluctuations as well as potentially seed large scale structure. Our scenario differs from the standard paradigm for the generation of density fluctuations in the following way: Unlike the standard case where curvature perturbations for a given scale $k$ are generated at one time and are constant on superhorizon scales, in our case inflaton decay continues to source the perturbations through the end of reheating, leading to growth of perturbations even on superhorizon scales. Here, we have considered two cases: (i) the case of a single-field inflation model in which the inflaton (not a SM field) decays into SM particles coupled through a Yukawa interaction and (ii) the effects of a non-zero Higgs effective VEV on the non-perturbative inflaton decay. For the case of non-perturbative decay, we considered an abelian gauge field coupling to the inflaton through a Chern-Simons term, as found in models of natural inflation~\cite{Freese:1990rb, Adams:1992bn, Freese:2004un}.

For perturbative inflaton decay to SM particles (with masses determined by the Higgs VEV), we find that, for high scale inflation with $m_\phi \sim H_I$, fermions with SM Yukawa couplings larger than $y \gtrsim 0.1$ would overproduce temperature fluctuations in the CMB (see Fig.~\ref{fig5}), unless one considers low values of the inflaton decay rate $\Gamma_0\lesssim 10^{-4}\, m_{\phi}$ and correspondingly lowered values of the reheat temperature. For reheating into the top quark, the reheat temperature must be lowered to a point where tension can arise with certain thermal leptogenesis models.  For scenarios in which the inflaton decays primarily to one of the majority of SM fermions with smaller Yukawa couplings, a variety of upper bounds can be set on the inflaton decay width similar to those indicated by Eq.~\eqref{eq:fittedfunc}. However, such constraints can vanish when reheating into the lightest SM fermions and can be significantly relaxed when the scale of inflation is lowered, as shown in Table~\ref{tab1}. In the case of parametric resonance, the effects of Higgs modulation lead to density perturbations that are significantly larger than those observed in the CMB, even when assuming relatively small couplings, $g = 0.1$, for SM gauge bosons at the inflation scale.  As a result, preheating into Higgsed gauge bosons cannot be the main source of reheating the Universe. If the inflaton couples to both a massless gauge boson (photon) and to massive ones ($W^\pm$ and $Z$), the problem becomes highly parameter dependent and is left for future work.

In summary, Higgs-modulated reheating can significantly constrain the parameter space for  inflationary models where reheating occurs by inflaton decay to SM particles. Even though quantum fluctuations of the inflaton may produce the observed spectrum of temperature anisotropies in the CMB, any realistic model must also provide for a mechanism to reheat the Universe. Typically considered as independent challenges, we have demonstrated that Higgs modulated reheating could potentially ruin the spectrum of density perturbations produced by quantum fluctuations of the inflaton. We note that specific models of inflation which reheat preferentially into either photons or the lightest SM fermion species are unaffected.  Without a concrete inflationary model which is demonstrably able to avoid the series of constraints we have calculated under a set of relatively generic assumptions, the most straightforward way to avoid the effects of Higgs modulated reheating is to introduce additional dynamics into the Higgs sector which stabilize its quantum fluctuations during inflation. Thus, Higgs modulated reheating can also be used as a window into the dynamics of the Higgs during inflation and, potentially, as a probe into physics beyond the SM.

\begin{acknowledgments}
We would like to thank Richard Easther, Sarah Shandera, and Spyros Sypsas for useful comments. AL, KF, and PS acknowledge support by the Vetenskapsr\r{a}det (Swedish Research Council) through contract No. 638-2013-8993 and the Oskar Klein Centre for Cosmoparticle Physics. KF acknowledges support as the Jeff and Gail Kodosky Endowed Chair in Physics at the University of Texas, Austin.  KF acknowledges support from the Department of Energy through DoE grant DE-SC0007859 and the Leinweber Center for Theoretical Physics at the University of Michigan. EIS  acknowledges support from the Dutch Organisation for Scientific Research (NWO). EIS acknowledges the support of a fellowship from ``la Caixa” Foundation (ID 100010434) and from the European Union’s Horizon 2020 research and innovation programme under the Marie Skłodowska-Curie grant agreement No 847648. The fellowship code is LCF/BQ/PI20/11760021. LV~acknowledges support from the NWO Physics Vrij Programme ``The Hidden Universe of Weakly Interacting Particles'' with project number 680.92.18.03 (NWO Vrije Programma), which is (partly) financed by the Dutch Research Council (NWO), as well as support from the European Union's Horizon 2020 research and innovation programme under the Marie Sk{\l}odowska-Curie grant agreement No.~754496 (H2020-MSCA-COFUND-2016 FELLINI). KF and LV would like to thank Perimeter Institute, where this line of research was started, for hospitality (KF is supported by the Distinguished Visitors Research Chair Program). The work of PS is partially supported by the research grant ``The Dark Universe: A Synergic Multi-messenger Approach" number 2017X7X85K under the program PRIN 2017 funded by the Ministero dell'Istruzione, Universit{\`a} e della Ricerca (MIUR), and by the ``Hidden" European ITN project  (H2020-MSCA-ITN-2019//860881-HIDDeN).
\end{acknowledgments}

\appendix

\section{Higgs decay through resonant boson production}
\label{appI}

We review the resonant production mechanism of gauge bosons responsible for the decay of the Higgs condensate, as described in Sec.~\ref{sec:backreaction}. The dynamics of the Higgs field doublet coupled to the gauge field $W_\mu^a$ is described by the Lagrangian term
\begin{eqnarray}
	\label{eq:higgslagrangian}
	\mathcal L_{\Phi+W} = (D_\mu\Phi)^\dag D^\mu\Phi - V_H(\Phi) + \frac{1}{4}G_a^{\mu\nu}G^a_{\mu\nu},
\end{eqnarray}
where the covariant derivative of the Higgs to the gauge field is $D_\mu\Phi = \left(\partial_\mu - ig\tau^a W^a_\mu\right)\Phi$, $g$ is a coupling constant, $\tau^a$ is a set of generators for the gauge group, and the field strength is defined as $G^a_{\mu\nu} \equiv \partial_\mu W^a_\nu -  \partial_\nu W^a_\mu + g\epsilon^{abc}W^b_\mu W^c_\nu$.

During the (p)reheating stage, the gauge couplings of the Higgs can be neglected, giving rise to the expression for the dynamical evolution of the Higgs field in Eq.~\eqref{eq:dhdN}. In terms of the conformal time $\tau$ and setting $\varphi = \sqrt{\lambda}ah$, the evolution of the Higgs field reads
\begin{eqnarray}
	\label{eq:Higgs_conf}
	\varphi'' + \varphi^3 - \frac{a''}{a}\varphi = 0\,,
\end{eqnarray}
where a prime indicates a derivative with respect to conformal time. During reheating, the inflaton field $\phi$ behaves as a massive scalar field of energy density $\rho_\phi$, so that $a\propto \tau^2$ and $a'' = 4\pi G a^3\rho_\phi/3 \approx \,$const. Thus, the latter term in Eq.~\eqref{eq:Higgs_conf} is important in the first stages of the preheating, while the cubic term takes over at later stages. The Higgs field sources the resonant particle production of gauge bosons through the dynamics obtained from Eq.~\eqref{eq:higgslagrangian}, as~\cite{Enqvist:2013kaa}
\begin{eqnarray}
	\label{eq:gauge}
	\WW'' + \omega_k^2\,\WW = 0,
\end{eqnarray}
where $\omega_k^2 =k^2 + q_W\,\varphi^2 - a''/a$ with $q_W = g^2/(4\lambda)$ and $\WW = aW_k$ where $W_k$ is the Fourier transform of the transverse component of the gauge field in Eq.~\eqref{eq:higgslagrangian}.\footnote{We ignore the non-Abelian self-interactions of the gauge fields, which may change the Higgs condensate decay time somewhat, but should not drastically affect our overall results.}

The gauge bosons are then produced resonantly through the oscillations of the Higgs field appearing in the term $\omega_k$ in Eq.~\eqref{eq:gauge}. This is similar to the production of the massive modes from the resonant oscillations of the inflaton field discussed in Sec.~\ref{subsec:PbPendens}.

We solve Eqs.~\eqref{eq:Higgs_conf} and~\eqref{eq:gauge} numerically during reheating. We find that various resonant bands exist where the W-bosons are produced resonantly, for different values of $q_W$. The corresponding occupation number is then~\cite{Greene:1997fu}
\begin{eqnarray}
	\label{eq:numberdensityW}
	n_k = \frac{1}{2\omega_k} \left(\left| \dot\WW \right|^2 + \omega_k^2 \,\left| \WW \right|^2 \right) - \frac{1}{2}\,,
\end{eqnarray}
\newline
from which we calculate the effective Higgs mass term induced by W-bosons using an approximate expression for the expectation value $\langle W^2 \rangle$ as in Eq.~\eqref{eq:gaugebosonmass}.

\bibliography{HiggsPert}

\end{document}